\documentclass[9pt,twocolumn,twoside]{pnas-new}
\usepackage{multicol}
\usepackage{graphicx}
\usepackage{subfiles}
\usepackage{cancel}
\usepackage{pdfpages}
\usepackage{hyperref}
\usepackage{pdfpages}

\newcommand{\MyTitle}[1]{
\begin{centering}
   \Large \usefont{OT1}{phv}{b}{n} #1        
   \par \normalsize \normalfont
   \end{centering}
   }
   
   \newcommand{\MySubTitle}[1]{
\begin{centering}
   \large \usefont{OT1}{phv}{b}{n} #1        
   \par \normalsize \normalfont
   \end{centering}
   }

\templatetype{pnasresearcharticle} 
\setboolean{displaywatermark}{false}

\title{Active Foam: The Adaptive Mechanics of 2D Air-Liquid Foam under Cyclic Inflation}

\leadauthor{Kroo} 

\significancestatement{Volume oscillation as form of distributed activity has been observed in biological systems during self-assembly and development. Here we demonstrate the mechanical effects of this perturbation in a purely synthetic analog: air-liquid foam confined to a 2D plane. We show in experiments and simulations, how cyclic inflation and deflation of a bubble effects the surrounding foam structure. We describe how this structure locally adapts over several cycles, while simultaneously propagating a subtle long-range "swirl" signal in the confined material.}

\authordeclaration{The authors declare no competing interests.}
\correspondingauthor{\textsuperscript{*}E-mail: manup@stanford.edu}


\author[a]{Laurel A. Kroo}
\author[b]{Matthew Storm Bull}
\author[c*]{Manu Prakash}

\affil[a]{Stanford University, Department of Mechanical Engineering, USA}
\affil[b]{Stanford University, Department of Applied Physics, USA}
\affil[c]{Stanford University, Department of Bioengineering, USA}

\affil[*]{e-mail: manup@stanford.edu}
\begin{abstract}

Foam is a canonical example of disordered soft matter where local force balance leads to the competition of many metastable configurations. Here we present an experimental and theoretical framework for "active foam" where an individual voxel inflates and deflates periodically. We explore structural adaptations of this disordered material with respect to added activity. Periodic injection of local activity leads to a small number of irreversible and reversible T1 transitions throughout the foam. Regardless of the presence of T1 transitions, individual vertices will displace outwards and subsequently return back to their approximate original radial position; this radial displacement follows an inverse law. Surprisingly, each return trajectory does not retrace its outbound path but rather encloses a finite area, with either a CW or CCW direction - which we define as a local swirl. These swirls form coherent patterns spanning the entire scale of the material. Using a reduced order dynamical model, we demonstrate that swirl arises as a direct consequence of spatial disorder in the local micro-structure. Building a first principles model, we demonstrate that disorder and strain rate control a crossover between cooperation and competition between swirls in adjacent vertices. Over longer timescales, the region around the active voxel structurally adapts from a higher-energy metastable state to a lower energy state, indicative of a localized annealing process. Finally, we develop a statistical toy-model that evolves edge lengths based on a set of simple rules to explore how this class of materials adapts over time as a function of initial structure. Adding activity to foam couples structural disorder and adaptive dynamics to encourage the development of a new class of abiotic, cellularized active matter.

\end{abstract}
\begin{document}
\flushbottom
\maketitle


\noindent \textbf{Key points:} Volume expansion and contraction in 2D air-liquid foam causes multiple kinds of structural adaptation that could be used in synthetic systems for material training, long-range mechanical signalling and rheological sensing. 

\section*{Introduction} 

Unlike most continuum materials, foam is a discrete, confluent, structured material. Similar to granular or cellular matter, foam is sensitive: local disorder deeply influences the material response to perturbation.
The collective motion of gas-liquid foam is driven by local force balances between surface tension at film intersections (“vertices” in 2D) and the subtle pressure differences between bubbles (which governs the curvature of edges) \cite{cantat2013foams}, \cite{bikerman2013foams}. 
Dynamical interactions between voxels are many-body, inherently coupled, and highly frustrated. Unlike spring or fiber networks \cite{ronceray2016fiber}, foam does not directly involve edge elasticity, which makes foam substantially more fragile against mechanical perturbation, and predisposed to bubble rearrangements ("T1 transitions"). These constraints impose significant nonlinearity, often even at very small perturbation amplitudes. This non-linearity enables foam to transiently occupy metastable structural states \cite{Hilgenfeldt1, durian2}, similar to many glasses. 

Soft disordered matter (such as foam) represents an opportunity to experimentally study materials at the breakdown of continuum theories; which in foam occurs at much longer length scales than in conventional matter (where this usually occurs at atomic lengthscales) \cite{mitov_sensitive_2012, hecke_jamming_2009}. Such materials promise deep conceptual insights into how we can generate accurate predictions in systems where the local micro-configuration matters \cite{machineLearningSoftness}. The utility of these predictions goes beyond simply shaping disordered materials for new applications at non-continuum lengthscales and may encourage new methods for strategically employing disorder rather than discarding it in functional materials. 


Active discrete systems have remarkable properties, including the capacity for learning \cite{machineLearningSoftness,dillavou2021demonstration}, classification/reconstruction \cite{fang2016pattern}, adaptation \cite{kotikian2019untethered,kori2017accelerating,bull2021excitable}, action at a distance \cite{buzsaki2004neuronal} and regulation \cite{imayoshi2013oscillatory, kiselyov2009harmonic}. For example, within the domain of biological systems, it has been demonstrated that activity is capable of significantly altering the yield stress of cellularized materials, influencing tissue shape during embryo development and driving many different self-assembly processes \textit{in vivo} \cite{vivek,shahaf,mongera2018fluid,bull2021excitable}. However, significant complexity arises from biochemical feedback processes in such systems.  There is value in studying purely synthetic cellular analogs, to identify and isolate complex biochemical feedback from the purely mechanical phenomena that arise from activity.

Activity in cellularized materials is traditionally introduced in one of two ways. The first method is to inject energy into the system through a boundary. For example, Corte et al. (2008) \cite{pine} showed that self-organized non-fluctuating quiescent states were an emergent phenomena in periodically sheared particle suspensions. Many others \cite{mukherji2019strength,teich2021crystalline,wang2022propagating} have recently contributed to a rapidly developing field, studying the role of oscillatory boundary perturbations on a variety of different types of granular matter and demonstrating that discrete, cellular-like materials exhibit rich dynamical behavior. Considering foam specifically within this class of discrete materials, Park and Durian (1994) \cite{durian1} showed that at the scale of large expansions (where the expanding cavity is much larger than the mean voxel size),  foam will develop fingering instabilities in response to a radial boundary perturbation, where the spatial periodicity of the instability pattern is driven by the strain-rate of expansion \cite{durian2}. This points to non-trivial dynamical behavior in foam, even in the bulk phase. 

The second method to introduce activity into these systems is to inject energy internally, at the same lengthscale as the system microstructure itself. This allows for a larger number of degrees of freedom, and allows the activity to be spatially distributed. In the context of using such internal activity to train materials, recent work \cite{nelson2017design, Berthier,hexner2020periodic,goodrich2021designing,dillavou2021demonstration,hagh2021freethenfreeze,kim2021embryonic} has explored how internal perturbations impart material adaptation. For example, Hexner et al. (2020) \cite{hexner2020periodic}, shows that periodic shortening and lengthening of specific edges in vertex models may be used to train materials to have auxetic and other unusual mechanical properties. Tetley et al. (2019) and Kim et al. (2021) showed that local, noise-like variations in edge tension resulted in control of material fluidization, and used this model to study wound healing in Drosophila tissue \cite{tetley2019tissue} and shape control in the development of zebrafish embryonic tissue \cite{kim2021embryonic}.  These recent studies are examples of active systems that are internally driven, which explicitly demonstrate non-trivial functionality. While highly informative, a current limitation of these systems is that experimental prototypes at scale are difficult to build. Because of this, the field has had sparse direct experimental support, especially in fully-synthetic (non-biological) systems.    

\begin{figure}[ht!]
\centering
\includegraphics[width=\linewidth]{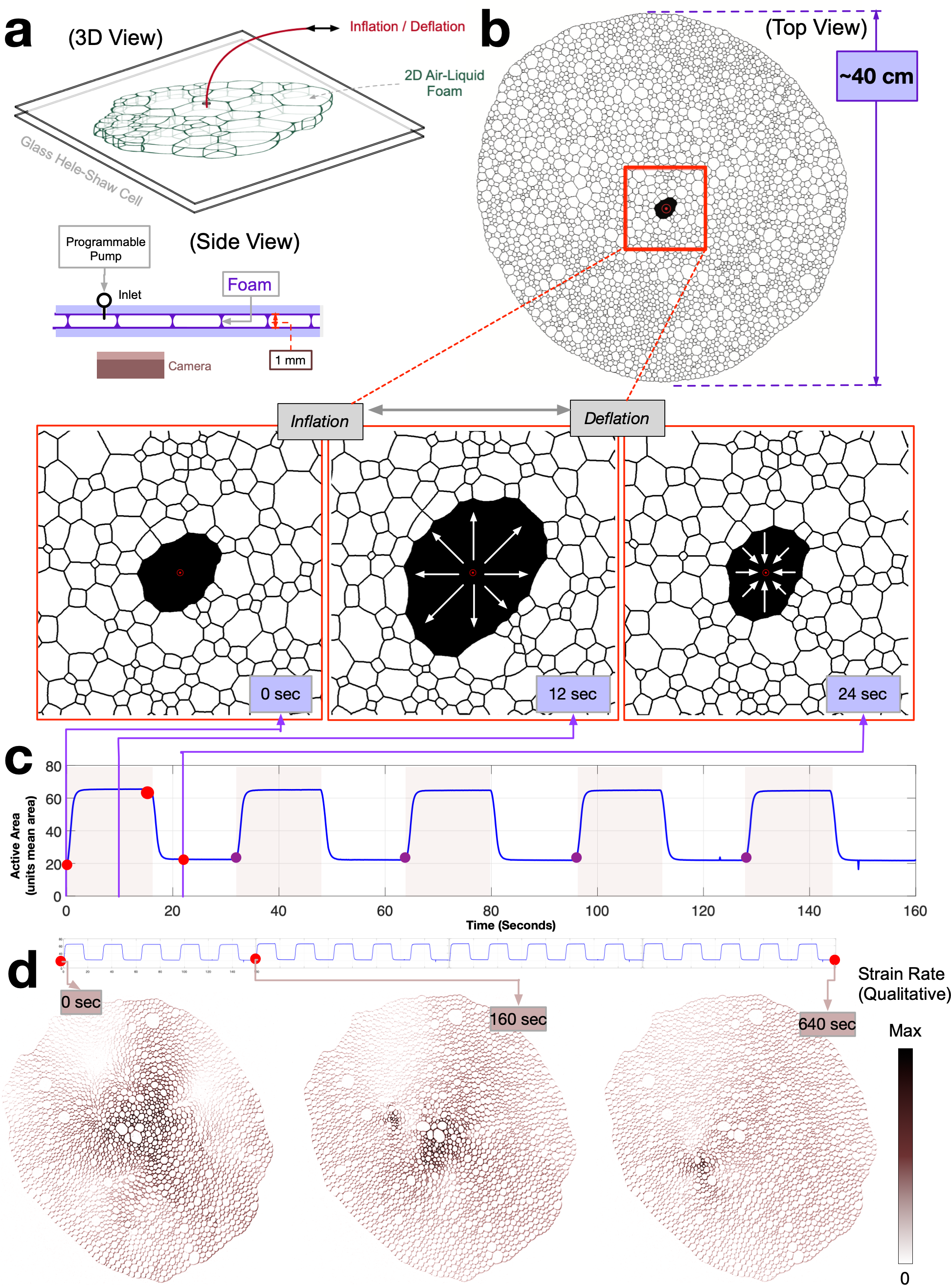}
\caption{ Experimental setup for active foam. a) Our experimental platform for studying active foam consists of a single layer of soap-air bubbles between 2 glass plates (in a hele-shaw cell). One or more small holes (significantly smaller than bubble size) in the upper plate allows for a programmable syringe pump to inflate or deflate active bubbles within the foam sheet. As seen in the side view, both the top and bottom plates are wetted with surfactant to prevent contact line pinning.
b) The foam structures are large, approximately 40 cm across, typically consisting of O($10^3 - 10^4$) bubbles per experiment. The foam is relatively dry, with a volume fraction estimated less than 85\%. We examine the effect of slow, cyclic inflation and deflation on the surrounding structure.  
c) The foam is cyclically driven at a constant frequency with a programmable syringe pump. The signal is a ramp and hold, followed by a symmetric deramp and hold.
d) The structure adapts to this imposed internal activity through bubble rearrangements and small, accumulated displacements between cycles. Here we picture strain rate, through time-averaging a stroboscopic series of frames (a forward average of first frame of 5 successive cycles). }

\label{experimentalPlatform}
\end{figure}

In light of this, here we specifically focus on the response of 2D foam to internal, cyclic inflation and deflation of a voxel. Our goal here is to explore how cyclic volume oscillation at a given site effects the surrounding passive material, and to inform the use of this perturbation as a basis unit of activity (e.g. akin to particle propulsion \cite{takatori2014swim,fily2012athermal,levine2000self} in more traditional active matter).

The mechanics of expansion and contraction of an individual voxel allow perturbations to be locally controlled and spatially distributed within the material. Internal, localized perturbations have been shown to give non-trivial insight into the nature of different bulk materials.  For example, internal perturbations of embedded droplets \cite{campas2014quantifying} and bubbles \cite{ saint2020bubble} have been shown to act directly as micro-rheological sensors in complex fluids, biological tissues and cross-linked hydrogels\cite{kim2020extreme}. In biological tissues, material states known to be influenced by local activity (specifically fluid-to-solid jamming transitions) have even been implicated in direct active control of tissue shape (elongation in embryo development)\cite{mongera2018fluid}. Local expansion and contraction specifically has been shown to modulate the mechanical properties in epithelial tissue, making it more robust to tissue fracture \cite{shahaf, prakash2021motility}. 

Here, we present a robust experimental platform implementing "active foam" via inflation and deflation of 2D foam (figure \ref{experimentalPlatform}, SI Movie 1). We introduce several toy models to demonstrate specific mechanisms underlying observed experimental phenomena.  We study the statistics of neighbor-swapping events ("T1" transitions) in the material via experiment, showing that there is a clear region of influence of a single active cell, with rare long-range T1 occurrence in the far field. Next, we investigate distinctive features of the displacement field: non-closing and looping vertex trajectories near to the activity site. We quantitatively characterize these phenomena as a function of cycle number and spatial location.  With a simple dynamical model we show that, even in the absence of T1 events, this looping or "swirl" in the vertex trajectory arises from disorder and dissipation. Additionally, experiments show evidence of large scale patterning in the sign of the swirl (e.g. coherent regions of CCW vs. CW motion), suggesting that subtle, mechanical effects propagate long range in the material. We conclude our study by discussing the structural adaptation of foam from a higher-energy state to a lower energy state, suggesting a localized annealing process. A a simple stochastic edge-length re-sampling model is used to qualitatively capture many of the statistical properties of energy adaptation and T1 transition distributions over time. Results suggest that internal cyclic activity in foam is a promising technique for the control and shaping of soft materials, demonstrating highly robust structural adaptation, in addition to long-range mechanical response. Here we endeavor to elucidate connections between the geometric structure and the dynamics of disordered systems under localized internal drive, with both experiment and simulation.

\section*{Experimental Platform and Constraints}
The experimental platform (Figure \ref{experimentalPlatform}a) we introduce consists of a monolayer of air-liquid foam (with average bubble diameter on the order of 5mm) between two glass plates in a Hele-Shaw configuration (spacing = 1mm). 

We begin each experiment with an unperturbed, fresh foam structure. This foam contains a large number of bubbles (order $10^3-10^4$) with significant variation in bubble sizes (section \ref{polydispersityHistogram}), with an overall volume fraction $\Phi_{2D} = \frac{V_{air}}{V_{liquid}} > .85$ ("dry foam" limit). The bubble solution we chose (50 percent "Beeboo" solution, 50 percent distilled water, see: SI section \ref{sampleprep}) is highly stable over timescales on the order of 48 hours — more than an order of magnitude greater than the timescales of rearrangement and relaxation studied in the experiments. An important consequence of this stability is that pressure-driven migration of gas across films does not greatly effect bubble size over the duration of the experiments (minimal coarsening). This means that the distribution of bubble volumes (section \ref{polydispersityHistogram}) is effectively invariant for the timescales of these experiments. In support of this, we also did not observe any T2 transitions in these experiments. 

In order to induce activity in the foam - we devise a method to inject air at a specific site within the foam (Figure \ref{experimentalPlatform}b).
 A small hole in the upper plate allows for a syringe pump to programmatically control the volume of air injected or withdrawn within a central "active" bubble. The amplitude of inflation and deflation is kept to the order of the average bubble diameter, enabling a localized injection of energy. Sample preparation is discussed in further detail in SI section \ref{sampleprep}.

To probe the foam systematically, we imposed several bounds on the injected energy. First, the volume of air injected is kept equal to the volume of air withdrawn in each cycle. As shown in the experimental data in figure \ref{experimentalPlatform}c, the cycle-to-cycle volume control is exceptionally precise (programmable syringe pump with a reproducibility of $<0.2$ percent).  The frequency of inflation and deflation is constant over time, as is the amplitude. And finally, the inflation/deflation is slow. The period is 32 seconds per cycle (\ref{experimentalPlatform}c). The purpose of this slow actuation was in an attempt to avoid significant inertial effects (capillary number approximately $< 10^{-4}$, where foam relaxation timescales vary from $~0.1-1$ seconds depending on the local structure). Relaxation in foam \cite{durian3} is multimodal, and tends to depend heavily on the details of the structure (not just the dissipation from confinement). Feature extraction was then performed to identify the locations of centroids, vertecies, edges and connectivity over time, as demonstrated in SI Movie 2 (and discussed further in SI section \ref{segmentation}). 

\section*{Displacement Fields suggest irreversibility and hysteresis in mechanical response}

\begin{figure*}[ht!]
\centering
\includegraphics[width=0.9\linewidth]{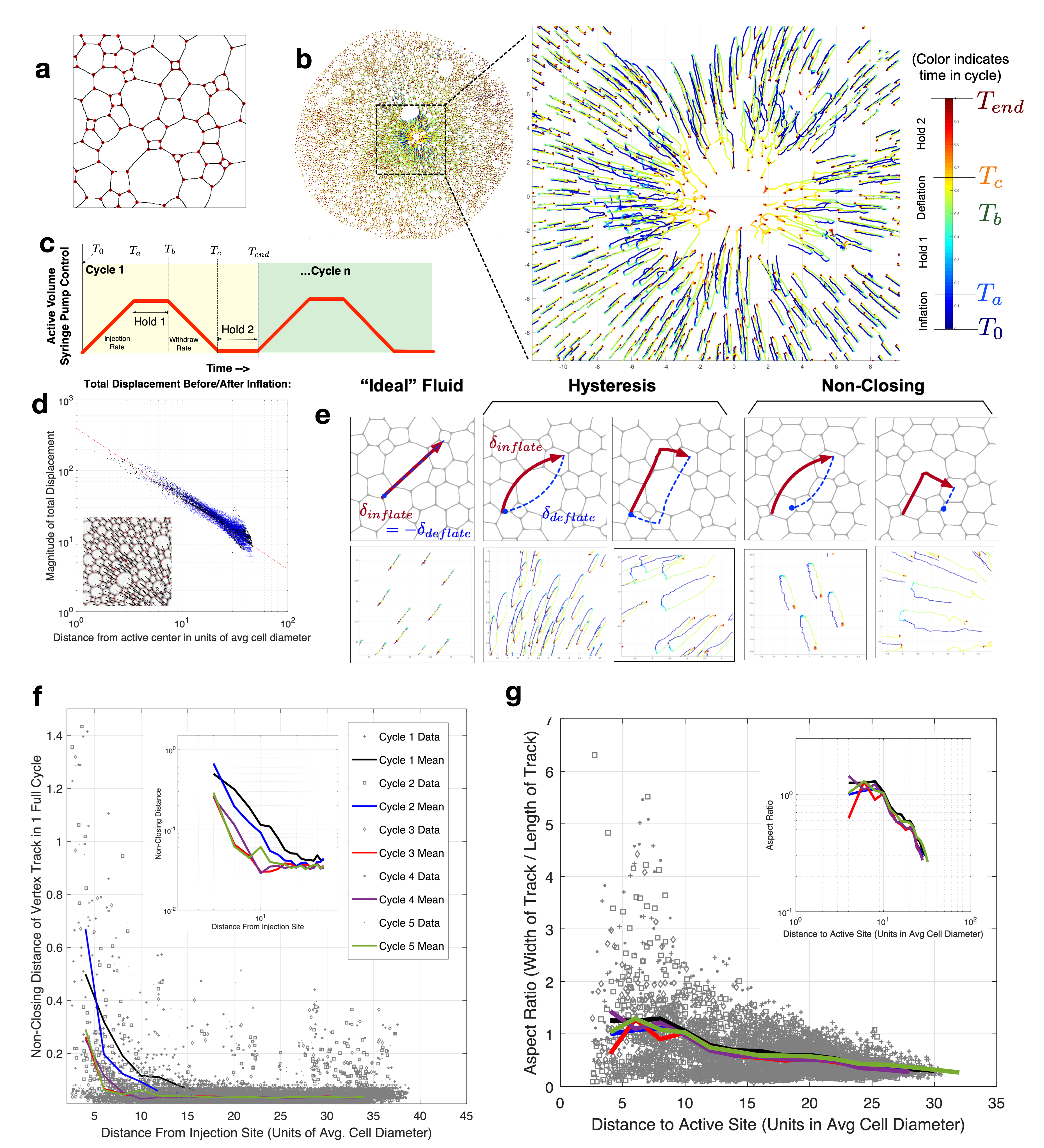}
\caption{Distinctive Features of the Displacement Field (a) Segmentation methods (described in SI \ref{segmentation}) allow us to identify and track the voxel centers, edges, and vertices in the foam. The pictured image is an example of experimental data after processing. Vertices are shown in red. 
(b) Vertex pathlines throughout a single inflation-deflation cycle are shown. Near the site of injection, the displacement field is complex. 
(c) The activity consists of a ramp and hold, followed by a symmetric de-ramp and hold. 
(d) The magnitude of these displacement falls off with 1/r away from the injection site. This is consistent with continuum fluids or elastic solids that obey continuity in two-dimensions (see SI sec \ref{radial}).
(e) Vertex pathlines exhibit three signatures: straight line paths, paths with hysteresis (enclosing an area), and non-closing paths. Hysteresis and non-closing trajectories appear to have multiple distinct shapes, suggesting multiple mechanisms at play. 
(f) Vertex Non-Closing Distance is shown as a function of distance from the active site. It appears to fall off approximately with 1/r, and successive cycles reduce the phenomenon of vertex-non-closing throughout the structure. 
(g) Hysteresis (particularly, 1/AR) is shown as a function of distance to the active site. Unlike non-closing, this phenomenon appears more persistent, less substantially influenced by successive cycling. }
\label{displacementField}
\end{figure*}

To understand the mechanical response of this air-liquid foam, we investigated the displacement of the vertex locations (plateau borders) and the bubble centroids, as shown in figures \ref{experimentalPlatform}d and \ref{displacementField}. Spatially, the net displacement of both centroids and vertices are inversely proportional to the distance from the site of injection.  This inverse-proportional response to inflation is what we expect from continuum systems regardless of if they are fluid-like or elastic (Fig. \ref{displacementField}d (details in SI section \ref{radial})). 

However, looking more closely at the displacement field, $u(r,\theta,t)$, depicted in figure \ref{displacementField}b, the displacement tracks of the vertices also have a significant non-radial component (figure \ref{displacementField}e). 

This is where the foam distinctly deviates from an ideal fluid or an elastic material: where path-lines of the material appear to have substantial tangential displacements, and non-closing trajectories after a full inflation/deflation cycle. If we characterize the area enclosed by these displacement-field traces, we can see that they frequently form “loops” of non-zero swept area. With regards to the non-closing distance, we will define this metric as the difference between a vertex’s position at the first frame of a cycle (before injection) and the vertex’s position on the last frame of the cycle (after deflation). This metric shows that subsequent cycles are driving the material to more consistently close tracks (figure \ref{displacementField}f-g). This phenomena (the lack of stroboscopic closure) is important, as it could be exploited to move the material over time, without changing the constraint set. These long-term non-reversible displacements can be visualized via the evolution of the strain rate fields in SI Movie 3 (and fig \ref{experimentalPlatform}d), for an example foam undergoing several hundred cycles in inflation and deflation (visualization tool:\cite{gilpin2017flowtrace}).

\section*{Structural Disorder causes Vertex Hysteresis}
In an effort to understand the mechanical origins of the phenomenological ‘loop’ nature of the vertex trajectories proximal to the site of injection, we develop a simple model that demonstrates that looping trajectories arise from the asymmetry of the micro-structure, where inflation and deflation are non-reversible operations. This structural asymmetry naturally gives rise to a discrepancy in relaxation timescales between vertices and edges as shown in Supplementary section \ref{simulationDetails}. 

\begin{figure*}[ht!]
\centering
\includegraphics[width=0.9\linewidth]{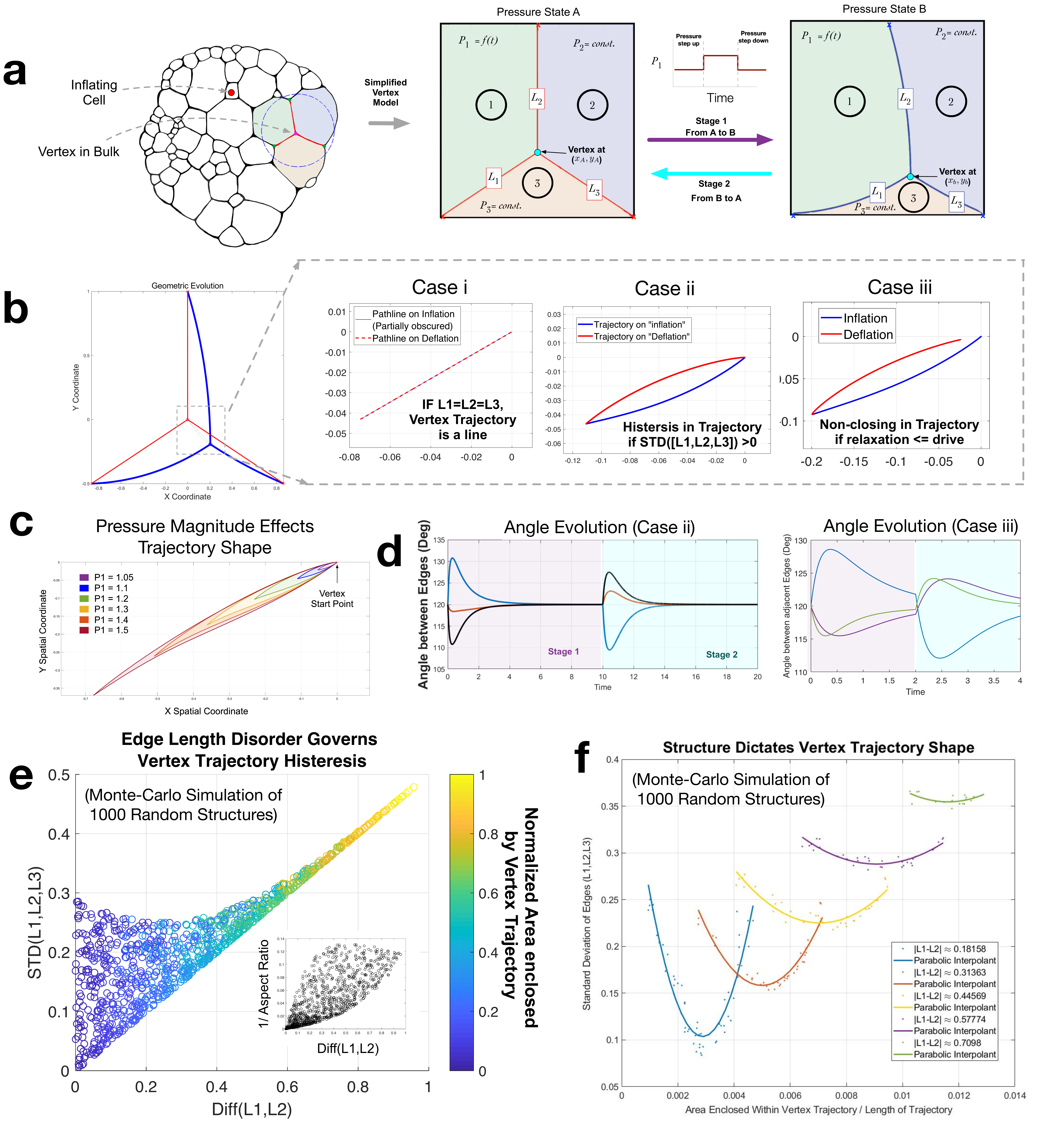}
\caption {A simple dynamical model for vertex response a) We constructed a simplified model to investigate the behavior of an arbitrary vertex position in the bulk of a foam structure experiencing cyclic activity. Considering a 3-bubble system with  ground state (pressure P1=P2=P3) - we subject the bubbles to a step function in pressure and allow it to relax. The step is then reversed and the structure returns back to ground state (details in S6). The pressure step is an analogy to the pressure signal transmitted throughout the structure due to distant activity.
b) By tracking the position of the vertex through time, we identify three cases: Case i, where all edge lengths are equal and the vertex trajectory is simply a line back and forth. Case 2 with disorder leading to a looping trajectory. Case 3, where the structure is not allowed to fully relax to equilibrium before the de-ramp, with evidence of non-closing trajectories (net displacement). 
c) The magnitude of the pressure step effects the trajectory shape and area enclosed under the curve. d) We can computationally assess the interior angles at the vertex as the shape evolves in time. Note that in case iii, the vertex angles do not reach 120 degrees before the pressure reverses. 
e,f) A monte-carlo simulation was performed to assess the relationship between disorder and hysteresis due to this “microstructural” mechanism (1000 structures were randomly generated). Note the relationship, where larger disorder in the structure tends to lead to a larger enclosed area of the vertex trajectory. }
\label{loopingVertexModel}
\end{figure*}

Focusing on a small region of foam, we build a three edge model (depicted in figure \ref{loopingVertexModel}a) with a few approximations. The first of which is that each edge in the system is, at all times, a circular arc segment. The arc segment is fully defined by a single degree of freedom, (defined as $h$), the height of the arc segment relative to the chord at the center-point between the two vertices. In addition to each edge having a single degree of freedom, each vertex has 2 degrees of freedom (position in x and y). For this small system, we studied a single mobile vertex attached to 3 edges — a system of five 2nd order, coupled nonlinear ODEs. This system comprises three edges, each with a single degree of freedom ($h_{1 - 3}$), and one vertex with two spatial degrees of freedom (in $x$ and $y$) as shown in Figure \ref{loopingVertexModel}a and further described in supplementary section \ref{simulationDetails}.

Each coupled ODE is given by a simple force balance which naturally gives rise to the equilibrium laws of foam: Plateau’s law (120 degree angles at each vertex) and Laplace’s law (curvature of edges is set by pressure difference between adjacent bubbles) - both of which are only strictly true at equilibrium. Unit testing was performed on a single edge to verify these fixed points hold true at equilibrium.

To compute the dynamics of the vertex, we calculate the force balance, given by:
\begin{equation}
   \vec F_{vertex} = \vec F_{\lambda,1}+ \vec F_{\lambda,2}+ \vec F_{\lambda,3}+ \vec F_{D,vertex}
\end{equation}

 $\vec F_{\lambda,1}$,$\vec F_{\lambda,2}$ and $\vec F_{\lambda,3}$ are constant, equal forces with a magnitude of $\lambda$ (total surface tension from both sides of the film), and a variable direction exactly tangent to the curved edge at the position of the vertex (as shown visually in SI fig \ref{freebodydiagram}) . The orientation of these three forces is defined by the edge curvature through this tangent constraint, coupling the vertex motion to the edge motion. $\vec F_{D,vertex}$ is the damping force due to vertex drag on the upper and lower plates. This force allows the vertex to dissipate energy in the form of heat as the vertex structure moves laterally in the Hele-Shaw cell. 

Next, we model foam edges as circular arc segments. This edge model was inspired by the work of Wesson et al. \cite{wesson2020steiner}, where abstraction of a surface as a simple, analytically-closed curve is used in place of discretization. In Wesson et al.\cite{wesson2020steiner}, they use ellipses to model droplet surfaces. The force balance on the edge node in the center of the edge is then given by:
\begin{equation}
    \vec F_{edge} = (\vec P_{Diff}) c+ \vec F_{\lambda}+ \vec F_{D,edge}
\end{equation}
Where $\vec P_{Diff}$ is the pressure difference across the edge, $c$ is the chord length between the edges, $\vec F_{\lambda}$ is the force normal to chord from surface tension, and $\vec F_{D,edge}$ is a nonlinear damping term for the edge curvature, proportional to the velocity to the $2/3$ power, as suggested by \cite{bretherton1961motion,kern2004two}. One could also classify this edge model as an example of a free-boundary height function method \cite{hirt1981volume}, carefully coupled to other degrees of freedom in the system. See details in supplementary section \ref{simulationDetails}.  
The boundaries of the system are fixed / pinned in this study (figure \ref{loopingVertexModel}a), such that each edge has a distal vertex that cannot move its position, and the pressures in the 3 cells are inputs to the simulation.

In this toy model we can study the motion of the mobile vertex in response to a step function in pressure of one of the 3 surrounding bubbles. The total force on each node is used to simultaneously solve the second-order system numerically, using a Fourth Order Runge-Kutta method.

The purpose here of constructing a model based on the dynamics of vertecies and edges is to uncover underlying mechanisms that drive micro-structural dynamics (e.g., pressure, surface tension, drag). 

The results of this model are shown in figure \ref{loopingVertexModel}b,c. The path line that a vertex takes in response to a step pressure perturbation is directly dependant on the three initial edge lengths. More disorder in the set of edge lengths leads statistically to a larger enclosed area by the path line (figure \ref{loopingVertexModel} e-f). If the pressure signal is reversed before the vertex has had time to fully relax to its equilibrium state, non-closing trajectories are possible, as demonstrated in 3biii and 3diii. Local edge disorder leads to hysteresis and non-closing trajectories. This points to a direct connection between structure and dynamics in our system, even without the presence of T1 transitions. 

We chose not to use a gradient-decent analog of this model \cite{brakke1992surface}, because it somewhat obfuscates the underlying forces involved (specifically, the explicit/quantitative tracking of pressure over time) by enforcing a cost function based on volume conservation. However, it is noted that energy-based minimization methods should (in principle) also be capable of demonstrating such phenomena, if the motion of the vertices and the edges are allowed different relaxation timescales. In other words, the looping and non-closure phenomena we demonstrate with this model do not appear to be inherently inertial in origin, as the behavior is extremely robust to large variations in the "mass" terms.

\section*{Topological Adaptation through T1 Events}

\begin{figure*}[h]
\centering
\includegraphics[width=0.9\linewidth]{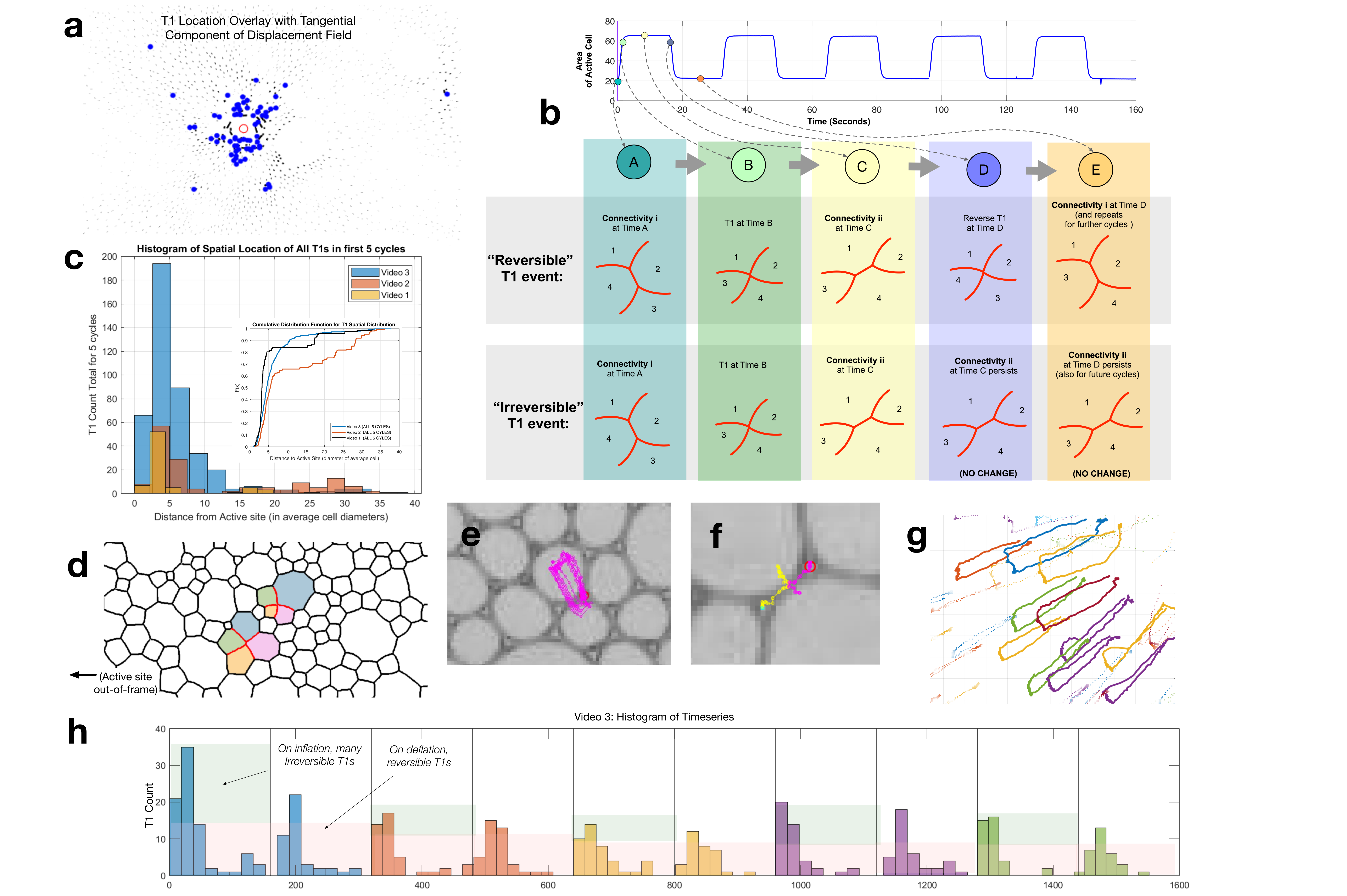}
\caption{ T1 transitions triggered by activity
a) T1s are microscopic yielding events that occur in the material in response to perturbation. Here we depict the tangential-component of the displacement field (non-radial), overlaid with the positions of these events (indicated in blue) for an example dataset. Notably, sparse T1s exist even in the far-field. 
b) We see evidence of two types of T1s in active foam: reversible and irreversible. Reversible T1s flip into one connectivity state upon inflation and then back to the original state upon deflation. Irreversible T1s cause permanent adaptation of the network. 
c) The spatial distribution of T1s (cumulative over the first 5 cycles) is shown for three example videos, demonstrating a proximal region where the T1s are common, and a long tail in the far-field. 
d) Each event always involves exactly 4 bubbles, where two bubbles gain a neighbor and two loose a neighbor. An example of two T1 events are shown here in this experimental data, occurring in close proximity to eachother within the same video frame.
e) Reversible T1 events tend to generate structures with a square-shaped hysteresis that match our observations in figure 2eiii. 
f) Irreversible T1 events tend to generate non-closing structures with an L-shape. The trajectories of two vertecies directly involved in an irreversible T1 are shown here. 
g) The “square-like” trajectories generated by reversible T1s are very similar to those observed in the data, in regions where T1s occur frequently (less than 10 avg. cell diameters from the injection site).
h) A histogram is shown of number of T1s detected vs. time over 5 successive inflation-deflation cycles. Note that there tend to be more T1s upon inflation than upon deflation (the difference accounting for the number of irreversible events in that cycle). These observations are consistent with the data in figure 2, showing that hysteresis tends to be persistent, whereas non-closing will tend to diminish over many cycles.}
\label{T1}
\end{figure*}

We perform computational segmentation and vertex tracking on over $10^4$ image frames and T1 identification of approximately order $10^3$ T1 transitions. Segmentation and T1 identification codes are discussed further in the supplementary sections \ref{segmentation} and \ref{T1Identify}, respectively. As with the displacement statistics, we observe a region near the activity that is most heavily effected by the T1s (figure \ref{T1}c). This region is weakly dependent on the injection amplitude and the polydispersity of the initial foam. This region is typically on the order of 4-8 average cell diameters away from the site of activity. This is because the nature of the input drive is geometrically forcing these transitions \cite{khan1986rheology} to occur close to each other, due to the 1/r scaling of the radial displacement field. 

T1s have long been of interest because they represent a form of microscopic yielding. The presence of such events can imply a soft "fluidization" state in materials \cite{bi2015density}. Typically in soft matter, these events are relatively rare and spatially distributed - they have been shown in simulations to have a measurable effect on the local structural energy \cite{machineLearningSoftness}.   

As shown in SI Movie 4 and Fig \ref{T1}b, we see that there are two distinct classes of T1s in this particular system: some T1s are “reversible” in that they flip back to their original position upon deflation in each successive cycle. These events create rectangular-like trajectories in their vicinity (figure \ref{T1}e).In contrast, many other T1s do not return to their original configuration. These “irreversible” events are permanent topological changes in the structure that contribute to the material’s long term structural adaptation. These events result in open, L-bracket path lines shown in figure \ref{T1}f. As shown in figure \ref{T1}h, we can track the occurrence of these reversible and irreversible events over the entire life of the experiment. The irreversible events tend to occur predominantly within the first 5-10 cycles. This disappearance of irreversible T1s is evidence of an annealing process as we cyclically drive the system towards a lower energy state.

Reversible T1 events tend to occur predominantly very near to the site of injection. However, the existence of irreversible T1 events (especially those in the far field) is surprising, based on the slow timescales of inflation relative to T1s, and the prior simulation-based studies.  Indeed, the ability of the bubbles in the foam to deform and reorient was shown in Cox et al. \cite{cox2008screening} to cause a strong “shielding” effect. We believe this unexpected experimental observation of irreversible T1s occurring in the far-field is due the strong effects of confinement in this experimental system, which was previously not accounted for. 

These metrics support the idea that there are several signatures of foam dynamics close to regions of activity. This  cumulative effect of T1s transitions and increased dynamical activity, creates a “region of influence” that the activity has on the foam. Results indicate that this region of influence is dependent on the initial foam structure, the magnitude of the inflation/deflation, and also the number of cycles driven. Next, we explore long range effects of activity on the dynamics of foam far from the active site. 

\section*{Long-Range Clockwise or Counter-Clockwise Signals}
\label{chiralitySection}
\begin{figure*}[ht!]
\centering
\includegraphics[width=\linewidth]{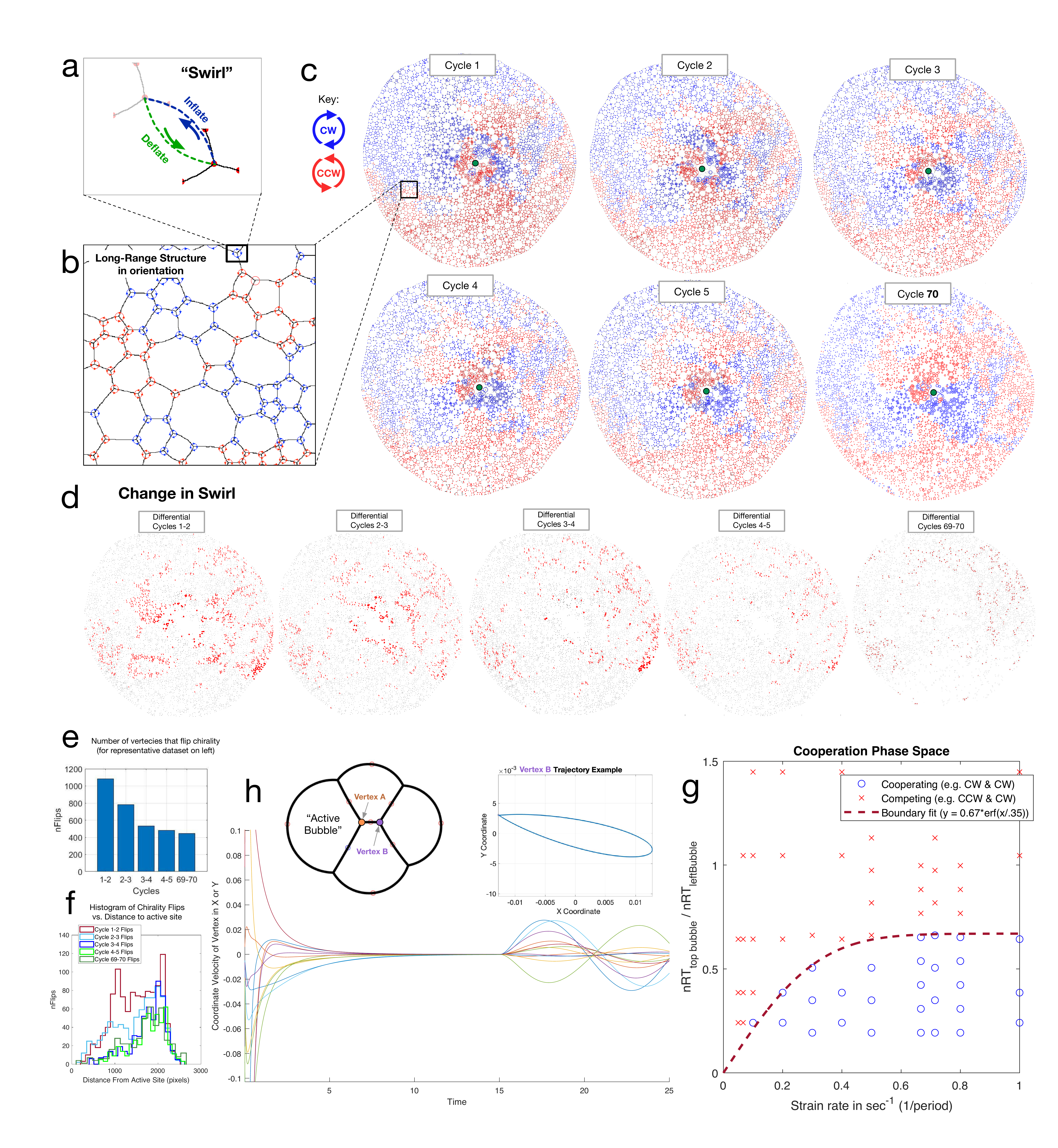}
\caption{ Swirl dynamics (a) For all nodes with looping trajectories during an inflation/deflation cycle, a "swirl" can be defined at every point in the foam as a CW (blue) or CCW(red) state. Computing this "swirl" for all vertecies, as depicted in (b,c) shows grain boundaries between cohesive CW and CCW swirl regions. Evolution of this pattern is depicted in (c) over first 5 cycles. Although the grain boundaries between CW/CCW regions shift over time, the pattern is persistent over many cycles (70th cycle represented in last panel in c). We observe that these coherent regions of cooperation decrease in size over time, resulting in a more 'mixed' swirl state. These coherent regions suggest some level of cooperation between adjacent vertecies.  (d) The cycle-to-cycle change in the swirl over time appears to converge within the first few cycles, arriving at a steady state after 5-10 cycles, as depicted in (e). This diminishing effect appears to have a strong spatial dependence (f), where after a few cycles, swirl flips only occur far from the injection site. (h) To further explore cooperative dynamics between a pair of adjacent vertices, we use a force-based numerical model of four bubbles. While one bubble on far left is inflated/defalted - we track the two center vertices, labelled Vertex A (orange) and Vertex B (purple) and identify the corresponding swirl (CW or CCW) of each vertex. The plot shows (arbitrary colors) the twelve x and y velocity time-series of every vertex in the 4-bubble system during first a relaxation stage from t=0 to t=15 sec, followed by a single pumping cycle, shown here with a period of 10 seconds. (g) In order to identify key parameters that control this cooperative dynamics in swirl, we draw a phase space systematically exploring structural disorder and variation in strain rates (rate of inflation/deflation). A clear separation between cooperative and competing dynamics is visible, with high structural disorder (e.g. disorder increases as $\frac{nRT_{\text{top}}}{nRT_{\text{left}}}$ diverges from $1$ to $<<1$) leading to cooperating dynamics.}

\label{chirality}
\end{figure*}

As demonstrated in figure \ref{loopingVertexModel}, hysteresis in the motion of vertices is evident throughout the foam due to structural disorder. However, upon further inspection, we observe that in addition to disorder influencing the presence of this effect (non-zero enclosed-area), vertecies influence each other locally, conspiring to form coherent regions within the foam (Fig \ref{chirality}b).  These large scale patterns form apparent grain boundaries between regions of CW (+) or CCW (-) swirl. We are careful here to use the term "swirl" rather than "chirality", as the trajectory direction is not invariant when superposed upon its mirror (which is the strict topological definition of "chirality"). 

Robust measurement of "swirl" direction (+ or -) is achieved by using a mathematical analog to Lock-In Amplification from the field of signal processing (see: Supplementary Section \ref{lockin}). The result of this method is an exceptionally sensitive measurement of the CW/CCW (+/-) direction of each vertex in time, as shown in figure \ref{chirality}a-b. 

We perform this analysis on 15 datasets, during the first 100 cycles of pumping. The series for video 1 for the first 5 cycle datasets is depicted in figure \ref{chirality}c, where the color corresponds to the CW/CCW (+/-) direction, and the size of the circles is proportional to the approximate area enclosed by the trajectory. 

As the foam is cycled, long-range swirl signals  emerge, propagating from the site of activity all the way to the boundaries of the foam. As the foam is cycled successively, these large-scale structures appear to shift in a subtle manner. These shifts sometimes occur via slow creep of a 'grain boundary' (lines on the differential plots in figure \ref{chirality}d) or via a whole region flipping within a single discrete cycle (block regions in figure \ref{chirality}d). The latter suggests that some of these long-range mechanical signaling dynamics may involve non-metric connections; where the dynamics arise from spontaneous symmetry breaking rather than from local pair-wise or multi-body interactions. This observation is consistent with prior literature \cite{mandal2020extreme}, which suggest similar such phenomena that may occur at especially high densities of active matter.  

Over time, fewer nodes change their swirl direction; for example, cycle 70 in figure \ref{chirality}c is nearly identical in structure to cycle 5. This indicates adaptation of the structure towards a dynamical steady-state. However, if we consider many such foams, it is clear that the coherent structures of CW vs. CCW directionality appear to diminish in size, apparently 'mixing' the system over the course of 5-10 cycles. 

Next, we examine the mechanism behind this emergent "swirl cooperation" lengthscale in a simplified 4-bubble dynamics simulation, depicted in figure \ref{chirality}h.  By driving this 4-bubble system while parameterizing disorder and dynamics (rate of inflation), we study cooperation of swirl (CW/CCW; +/-) in the motion between adjacent vertices (see: the two interior vertices marked 1 and 2 in figure \ref{chirality}h). An example of the simulation output is depicted in SI Movie 5. 

To probe the cooperation between these adjacent vertices, we systematically changed the initial volume of gas in the uppermost and leftmost voxel (whereas non-equal volumes for the bottom and rightmost bubbles were set and kept constant for all configurations). The strain rate of activity was also systematically varied (e.g. increasing the period of a single-cycled sine wave). As shown in figure \ref{chirality}h, each configuration was allowed to fully relax before a single inflation-deflation cycle was performed. The vertex trajectories were then categorized as "clockwise" or "counter-clockwise" as an output of the simulation. An example vertex trajectory for a single cycle is shown in figure \ref{chirality}h.

We find that cooperative dynamics (e.g. the tendency of two adjacent vertecies to swirl in the same direction) depend strongly on two key parameters: the initial disorder in the system (as a nontrivial function of all 4 bubbles volumes), and the strain rate of the pumping  (where a faster strain rate corresponds to more cooperative behavior). This points to a complex mechanical signal that is inherently dependent on both initial structural properties (e.g. static inputs) and also on purely dynamic parameters (pumping rate). The implication is that coherent swirl patterns report on both the local structure and perturbation dynamics.

Indeed, this intuition matches the spatial patterning we see in figure \ref{chirality}c, where the fastest strain rates will likely be near the injection site when T1s are occurring frequently in the first few cycles. In these regions, there appear to be larger regions of coherent swirl structures, consistent with the simulation results. Similarly, after many cycles, as the material becomes more ordered in the vicinity of the injection site (and when T1s occur less frequently), these coherent structures appear to decrease in characteristic size. 

This is further shown in figure \ref{chirality}d-f where we examine the number of vertecies that flipped from CW to CCW, or visa versa, within a single cycle. The adaption of vertecies in their directionality of swirl appears to decrease for the first 5 or so cycles, levelling off to approximately constant thereafter. As shown in figure \ref{chirality}f, this "levelling-off" is highly dependent on spatial position. The regions close to the injection site appear to have very minimal number of flips versus the regions far from the injection site remain predisoposed to scattered flips, even after many cycles. 

This spatial dependence is likely because of differences in the regional relaxation times. Close to the injection site, vertecies undergo vastly accelerated relaxation due to their proximity to activity. Because of this, proximal regions to the injection site quickly become "caged" in their dynamics (e.g. subdiffusive MSD in supplementary figure 9 after 5-10 pumping cycles). By contrast, the farfield remains in a state of slow but consistent change, resulting in scattered "flips" even after dozens of cycles of activity.  

To explore how swirl might be useful as a parameter in a range of disordered active systems, we compare our foam datasets to simple spring network models. If we replace the foam edges with springs, and lift the volume constraints of bubbles (simply a disordered spring network), it is clear that the long-range swirl dynamics are not reproduced (see supplementary section \ref{springs}) -- although some CW/CCW signal is still present due to the initial disorder. The strict constraints of bubble volumes, and top/bottom confinement are thus critically necessary factors for long-range coherence and structural propagation of the mechanical signal encoded in swirl dynamics. 

\section*{A reduced-order statistical model of edge-length dynamics}

\begin{figure*}[ht!]
\includegraphics[width=0.9\linewidth]{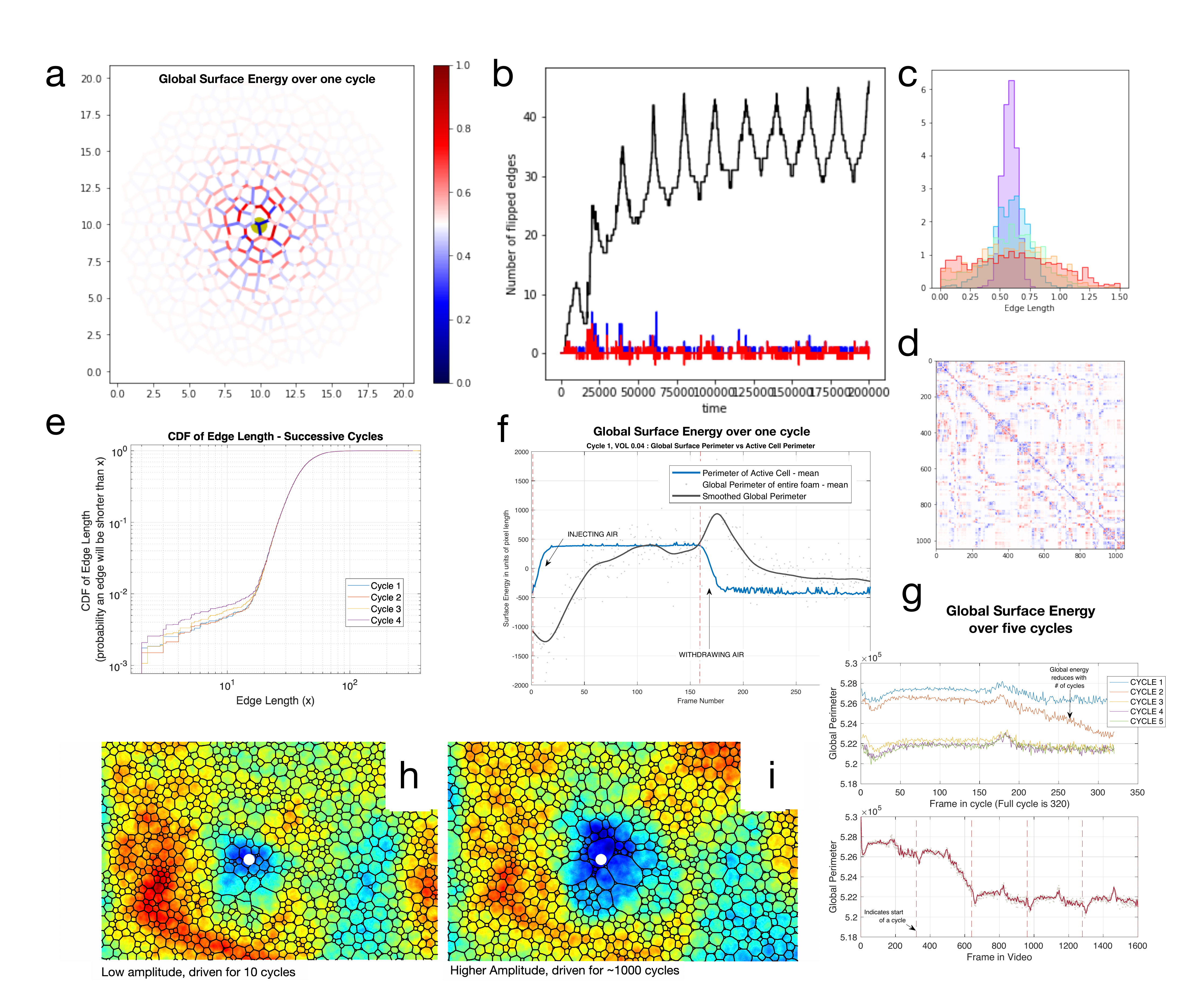}
\centering
\caption{ Statistical model and energy landscape (a) We present a statistical reduced order $0+1$ dimensional model of a foam-like network. An initial foam-like network structure shown here is generated using voronoi Einstein patterns. Each edge is assigned an initial susceptibility, depicted here with red as a positive susceptibility $ \chi_i $ (will tend to lengthen upon inflation), whereas those edges in blue have a negative susceptibility (tendency to shorten upon inflation). The opacity represents the magnitude  of susceptibility (darker regions have a stronger response). (b) By adding activity to this foam-like network (as discretized by the input activity function $dA$), we can track the number of flipped edges in the network as a function of time. The black line shows the cumulative number of flips over the course of 10 cycles (where flips reversing to their original state can contribute negatively). The red line shows the number of flips that occur with a positive susceptibility, where blue shows flips with a negative susceptibility. The black line qualitatively captures the same behavior we see in experimental T1 data for "active foam", figure \ref{T1}h. (c) We can experiment with this system by inputting different initial structural configurations (example edge length distribution histograms shown here) and watching these distributions reduce the number of short edges (d) The connectivity and strength of local coupling can be represented as a matrix $\eta_{i,j}$ (equation $6$) where the $x$ axis is the $i$th edge and the $y$ axis is the $j$th edge. Red vs. blue color represents the effect that shortening the $i$th edge has on the length of the $j$th edge.
(e) This reduction in number of short edges is also observed in our experiments. This edge-lengthening effect occurs even while polydispersity and coordination number stay constant (SI \ref{polydispersityHistogram}). 
(f) We provide a metric for energy stored in the surface of the foam by computing the sum of total lengths of all edges in the foam. This is completed for each frame, and the average is shown in black. The blue line shows the perimeter of the active cell. Note that at the beginning of both the inflation and deflation, the energy temporarily reaches a minima or maxima, respectively, while the structure is not in mechanical equilibrium. 
(g) The energy over 5 successive cycles is shown. Consistently we can see that the effect of cyclic inflation/deflation is an experimental example of an annealing process on a highly complex energy landscape. The structure appears to descend chaotically through a series of meta-stable states  (even though successive cycles are identical in amplitude and location, and the underlying dynamical system is deterministic). 
(h) To highlight the role of local adaptation of foam to activity - we can also compute the local energy by summing the lengths of edges within a certain radius of each pixel. Here, the color indicates energy (normalized from min to max). Note that after 10 cycles at a low amplitude, a region of low energy is written into the structure of the material proximal to the activity. 
(i) For a higher amplitude and for a larger number of cycles, we can see that the region of low energy in the structure can grow; suggesting the effect is tunable - thus supporting the idea of structural memory in foam and foam-like materials.}
\label{energy}
\end{figure*}

In our experimental work, we find that successive cycling of the foam tends to change the  distribution of edge-lengths (figure \ref{energy}f) over time. This process, where short edges are gradually annealed into longer ones, is important because the bubble volumes on average do not change, nor does the average coordination number. The structural adaptation is occurring predominantly along this dimension of edge-length, as the system rearranges while also obeying extremely stiff constraints (due to the low compressiblility of bubble volumes, and the top/bottom confinement of the glass plates). The minimization problem is so confined and stiff, that solving it with conventional methods (Surface Evolver \cite{brakke1992surface}, Viscous Froth Model \cite{kern2004two}) poses significant challenges with numerical stability and computational resources at large scale (> O($10^3$) bubbles), and currently remains difficult for many applications. Very recent advances \cite{karnakov2021computing} in multi-layer Volume of Fluid Methods are notable and encouraging. 

To probe the enormous phase space of possible injection amplitudes and initial foam structures, we developed a new reduced-order statistical model of the edge-length dynamics in 0+1 dimensions, without modelling their physical location or tracking their exact orientation. Critically, this model tries to generalize the essential, bare-bones features of a foam to a tractable high-dimensional, nonlinear dynamical system.

A foam is a metastable material in a disordered energy landscape, so in order to study it to leading order, we define an external perturbation analogous to our experimental inflation-deflation cycle. We capture these dynamics by mapping the stimulus onto a change in each edge length and define the edge 'susceptibility' of the $i$th edge, $\chi_i$, to each stimulus, $\delta A_k$. We note that this leading order susceptibility is not explicitly a linearization of the dynamics but instead exploits the idea that the start and end points for an edge-length at the start of the pumping and the end can be decomposed as a distance $\Delta L_i$ which we can then divide by $\delta A_k$ to extract the average susceptibility of the cell for a given $\delta A_k \in $ scalar (which is analogous to the mean value theorem). For each edge, We denote this response as the susceptibility: $\chi_i (\delta A_k)$.

This definition allows us to define the change in length of each edge in response to stimulus as:
\begin{equation}
    \mathcal{Y}(\delta A_k) \approx \vec\chi*\delta A_k
\end{equation}

Where $\delta A_k$ is positive upon inflation and becomes negative during deflation. From experimental measurements, we know that some susceptibilities will be positive and some negative, or in other words: some edges are predisposed to growth or shortening, in response to the same stimulus. 

Next, we introduce the nonlinearity of this model in the form of a characteristic feature of foams: edges undergo t1 transformations of their topology in response to stress. To allow for reversible t1s we use a simple rule when the $i$th edge reaches the t1 threshold (equals 0 in length) we simply flip the susceptibility to driving by a sign. 
\begin{equation}
    \chi_i^\star = -\chi_i
\end{equation}
This choice allows for perfectly reversible events as we cycle through $\delta A_k (t) \in \mathbb{R}$ with the minimal number of parameters.

The observation that the susceptibility of foam edges to stimulus can be both positive and negative suggests a perhaps useful analogy to other forms of re-configurable bonds seen throughout soft materials research \cite{mitov_sensitive_2012}: the concept of the 'slip' and 'catch' bonds \cite{rakshit_ideal_2012}. These two classes of bonds are defined by their phenomenological behavior. Slip bonds increase the probability of breaking with increased applied force. Whereas, catch bonds increase their stability and resilience in response to applied force \cite{kachman_self-organized_2017}. Drawing an analogy between T1s at $\ell_i = 0$ and bond rupture, suggests that our model of edge dynamics may behave similarly to an ensemble of catch and slip bonds in relative population. Those with $\chi_i < 0$ become more susceptible to bond breaking with increased stimulus (e.g. similar to a slip bond). While those edges where $\chi_i > 0$ become more stable with increasing stimulus (conceptually similar to catch bonds). Such an analogy highlights the essential intuition that the structural adaptation of foam is rich and nontrivial even in its simplest form.

Foam is known to encode a memory in its structure \cite{mukherji2019strength}. Each fixed topology of edges defines a basin of attraction to a state which minimizes the perimeter while respecting the foam cell area constraints, but when you change the topology through a t1, the foam may minimize to a different state. We encode the topological configuration of the foam in by keeping track of the number of times that an edge has undergone a T1 transition with a structure which call "flip" which will be incremented each time a flip T1 occurs at the ith edge.  

\begin{equation}
    \text{flip}_i= \text{flip}_i + 1
\end{equation}

To capture the collective dynamics, the final ingredient to this minimal model couples the dynamics of these edges with a leading order coupling matrix. This coupling matrix allows us to play forward the dynamics of this 0+1 dimension model with N coupled degrees of freedom. We can then use the susceptibility, stimulus amplitude, and the flip bit to track the change of length on an edge:
\begin{equation}
    L_j^{\star}= L_j + (-1)^{flip_i}\eta_{i,j} + \chi_j \delta A
\end{equation}

where $\eta_{i,j}$ is a quenched disorder matrix which tells you how much edge length a t1 of the $i$th edge alters the rest length of the $j$th edge. This matrix is encodes the connectivity of the edge to other edges, and it's sphere of influence (any given edge can be connected to more than it's nearest neighbors, through a tuning the structure of this matrix).The evolution of this dynamical system in response to cyclic stimulus defines the adaptive mechanics of this model foam. In some ways, this high-dimensional nonlinear system resembles the mathematics of an integrate and fire neural network but is more deeply related to two-state models in facilitated dynamics (e.g. 'transformation X cannot occur until after Y has occurred') \cite{mitov_sensitive_2012}. This observation may hint to the dynamical complexity of even simple foam-like models.

With the leading order dynamics established, we discuss the initialization of the models quenched disorder. To help constrain the possible matrix statistics we can derive $\eta$ and $\chi$ from an Einstein disordered solid \cite{kachman_self-organized_2017} using the initial orientation of edges and their distances to the center (see fig \ref{energy}a).
\begin{equation}
\chi_i = (\chi_{mean} - sin(2\Delta\theta_i))e^{-d_i}
\end{equation}
where $\Delta\theta_i$ is the angle difference between the point of injection and the edge mean. Constrained by the physics of a single bubble, we assume that the mean of the susceptibilities ($\chi_{mean}$) is constrained to equal $\sum_i \chi_i = \sqrt{\frac{A_o}{\pi}}$ which is the increase in the perimeter of the pumped bubble if it is not interacting with any neighbors (e.g. perimeter of circle with a change in the area). This constraint on the initial distribution allows us to tune the system response by choosing to pump a larger or smaller initial bubbles cyclically.

A similar metric can be constructed for the coupling between edges (fig. \ref{energy}d):
$$
\eta_{i,j} = - cos\left(2\Delta\theta_{i,j}\right)e^{-d_{i,j}}
$$

We implemented this model for a number of synthetic networks to ask the question: how does this model material transition from purely reversible transformations to structural adaptation through irreversible T1s? Figure \ref{energy}a depicts the initial network topology and the susceptibility by color (red indicated the edge is predisposed to shorten with an applied stimulus; blue indicates the edge is predisposed to lengthen). This was initiated based on the distance of the edge from the injection site, and the initial orientation of the edge relative to the injection site. The connectivity, $\eta$ controls the strength of influence of any given edge on another local edge (fig. \ref{energy}d). We can control the network's initial distribution of edge length (figure \ref{energy}c) as an input parameter.  

In figure \ref{energy}b, we depict the qualitative result of T1 flips as a result of cyclic pumping. It is notable that both irreversible and reversible T1 events occur in this model, and that the adaptation of T1s in the model (figure \ref{energy}i) looks remarkably similar to what we see in experiments (figure \ref{T1}h): with irreversible T1s becoming rare over the course of 5-10 cycles. This leading order dynamical model provides an analytically tractable viewpoint for studying conceptual questions at the core of irreversibility in externally driven sensitive matter with structural memory.

\section*{Energy Landscape and Localized Structural Annealing}
 As depicted in figure \ref{energy}, we can measure the local (h,i) and the global energy (f,g) as a function of time. Because 2D dry foams are governed by surface tension, the potential energy stored in the structure of the foam is proportional to the sum of the lengths of the foam edges. We can use this property to compute the global energy (by summing all edge lengths in the foam), or to compute the local energy (by summing only those edges that are within a certain radial distance r, from any given pixel). In figure \ref{energy}h-i, a radius of 3 times the average cell diameter was used to compute the normalized structural energy (shown in the color map). 
 
 We observe both local and global structural annealing in response to cyclic inflation and deflation. Short-term transients in energy at the time of inflation and deflation indicate the structure is temporarily out of mechanical equilibrium. The long term decent through energy landscape appears to be strongly dependant on initial conditions, and is exceptionally difficult to predict due to path-dependencies of the structural adaptation.The fact that this experimental system rapidly adapts quickly to reach a structural potential energy minima (in the vicinity of the activity) is consistent with recent simulation work on soft deforming jammed particles \cite{Berthier,hagh2021freethenfreeze}.

\section*{Discussion}

Understanding foam represents an engaging challenge which highlights the vital importance of microconfigurational details in forecasting material response. This intellectual pursuit forces us to reconcile high-dimensional quenched disorder with nonlinear dynamics under constraint in search of an emergent simplicity. In this work, we address this intellectual frontier by taking the viewpoint of foam volume actuation. The first step in the direction of active foam is to master the response of the foam to a single, simple local stimulus and discover the rules which map observed microconfiguration onto the observed spatiotemporal response. In this effort, we have discovered a number of characteristic phenomena that arise in response to volumetric bubble driving. 

The first surprising result was the emergence of a remarkably long range mechanical signal in the material. This is visible in the long-tail of the T1 distribution as a function of distance to the active site, and is apparent in the local structural energy of the material over several cycles. Long-range mechanical signaling is of interest to the active matter community as it can enable far more complex material capabilities (for example, considering action-at-a-distance in neuronal systems, or sparsely-driven excitable mechanics). 

The second interesting characteristic phenomena we observed was that the foam’s adaptation was mostly complete by 5-10 cycles. This was shown in the ratio of reversible to irreversible T1s as a function of cycle number, the mean-squared bubble displacement over time, and was also recapitulated with the edge-length toy model – suggesting that this perturbation robustly drives adaptation primarily within small number of pumping cycles. In the context of active matter, this sets boundaries on the temporal response of the material, and sets expectations for effective control strategies.  

A third phenomena that was observed was the clear delinearation between reversible versus irreversible T1 transitions (voxel neighbor swapping) in the material. In the context of cyclically driven colloids, this identification implies that certain microconfigurational locations can be predisposed to yield, but with the potential to reverse the effect upon deflation.  Technically, these large yielding events are occurring – but many do so in a manner that can immediately reverse.  By contrast, irreversible T1 events are partially responsible for long-term adaptation of the material. 

Another phenomena observed was the presence of CW versus CCW “swirl” in the trajectories of the vertices, when in the presence of disorder.  Each vertex has a predisposition to turn CW vs. CCW in the material based on the local structure and location of the active site. This effect was demonstrated to arise from local anisotropy and a mismatch in dissipation across the local structure. We are careful here to note that the vertex itself does not rotate in its orientation – simply translates.

Further, we have shown that the cooperation between swirl in adjacent nodes is a direct consequence of both the structural disorder and the strain rate, pointing to a non-trivial many-body system that governs the cooperation of swirl in a material. The number of vertices that change their swirl directionality per cycle appears to correlate qualitatively with local relaxation time in the material, and appears to act a long-range mechanical readout of material adaption.  The amount and direction of the nodal circulation is determined by the details of the stimulus location, amplitude, strain rate and local microconfiguration. We have shown through simulation that none of these details is negligible -- highlighting the rich complexity of the problem and the challenges this poses for control / reduced order predictive techniques.
We have shown that a foam adapts to a repeated volume stimulus in a fashion which minimizes structural energy in the foam’s configuration – both locally and globally. This property could be used to shape a foam through a sequence of stimuli, training the foam’s material response. Using memory of disordered materials and employing the use of transient, periodic activity to train such materials has been proposed by many in the past \cite{mukherji2019strength, Berthier, manning1, nelson2017design}, and still remains a vast and challenging research space for further investigation.
Finally, framing a material like this as a biphasic, competing population of slip vs. catch bonds allows us to discuss adaptation of this material as a continuous function of input parameters (such as local disorder, etc). This framework allows us to begin to predict statistical average responses of the material to potential activity locations based on a reduced order model. 

In conclusion, these experimental observations (supplemented in many cases by simple, toy-model simulations) highlight the exciting potential for the use of dissipative foam materials as a useful, computationally adaptable complex system – capable of local energy minimization and a fascinating non-linear dynamical response to internal cyclic activity. The sensitivity of foam to its environmental history is simultaneously a challenge and an opportunity. It is challenging to predict the precise response of the material given only a single snapshot. Yet, the information encoded in the material presents an opportunity to make context-aware, adaptive machines which more closely reflect the success of stable-yet-sensitive living matter. Here, we seek an emergent understanding of foam adaptation to an understudied form of driving in foam -- the cyclic inflation of foam cells as the foundation of 'active foam'.

\section*{Acknowledgements}
MP would like to thank the Moore Foundation, Keck Foundation, Pew Foundation, National Science Foundation Career Grant (1453190) and NSF CCC Grant (DBI-1548297) for supporting this work. We would like to thank our colleagues at APS, 2019 Adaptive Materials GRC, 2019 Soft Matter GRC and 2020 EUFOAM conferences who have given valuable early-stage feedback on this investigation. LK and MB gratefully acknowledge support by the National Science Foundation Graduate Research Fellowship (DGE-1147470) and the Stanford University BioX Fellows Program (MB).

\section*{Author Contributions}
LK,MB,MP designed the research. LK performed the experiments, wrote experimental code and analyzed the data. LK developed the dynamical simulation model with input from MB and MP. MB developed the statistical model with input from LK and MP. All authors wrote and edited the manuscript. 
\section*{References}
\bibliography{REFERENCES}

\clearpage
\newpage
\onecolumn

\begin{centering}
\MyTitle{\textbf{Methods and Supplementary Information }}
\vspace{20px}
\MyTitle{Active Foam: The Adaptive Mechanics of 2D \\
Air-Liquid Foam under Cyclic Inflation}
\vspace{20px}
\MySubTitle{Laurel A. Kroo, Matthew Storm Bull and Manu Prakash}
\vspace{20px}
\MySubTitle{SI Movie Files 1-5:}
\end{centering}

SI Movie Files for Preprint are here:
\url{https://drive.google.com/drive/folders/1CoslaMBrYyhfZqpAHL9Flx4pTufXtVnC?usp=sharing}

\tableofcontents
\clearpage

\section{Comparison between Foam and an Ideal Potential Flow Source/Sink}
\label{radial}
For the purpose of identifying interesting signatures in the foam, we can compare any point in the material to a simple potential flow with an unsteady boundary condition. The result is shown in figure 7; where the black lines represent the predicted trajectory from this analytical model. Deviations from this prediction indicate non-newtonian mechanisms at play. 

\begin{figure}[ht!]
\centering
\includegraphics[width=\linewidth]{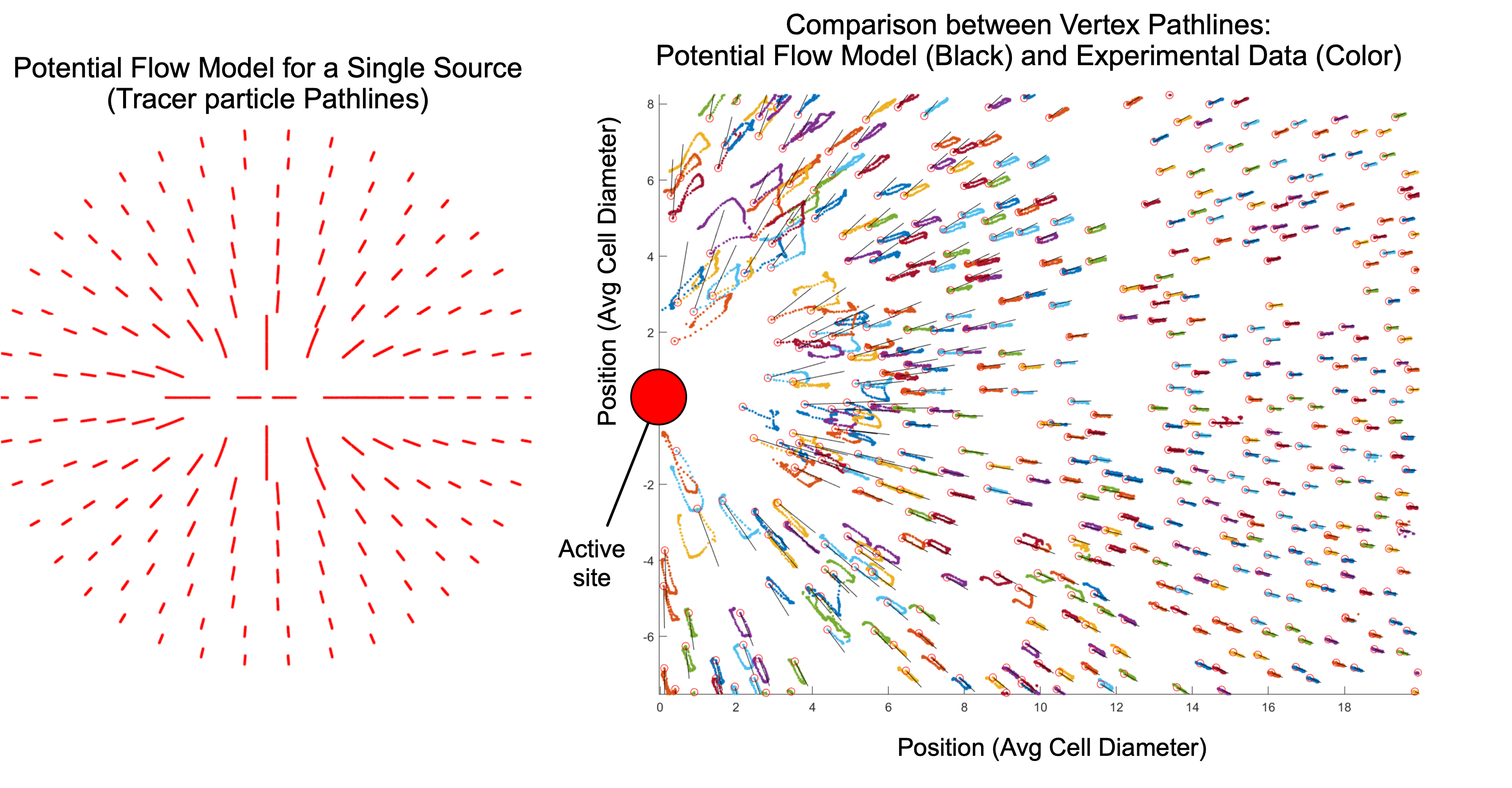}
\caption{On the left is the predicted tracer pathlines from an ideal potential flow. On the right is an overlayed version of the prediction (in black) with the actual experimental data in color. }
\label{fig}
\end{figure}

This analytical model is standard and is also cited in well-known texts in fluid mechanics (SI Ref. 5).
Consider a velocity potential in Cartesian coordinates (where $u$ and $v$ are components of \textbf{v}):

\begin{equation} \label{eq:1}
    u(x,y,t) = \frac{d\phi} {dx}
\end{equation}

\begin{equation} \label{eq:2}
    v(x,y,t) = \frac{d\phi} {dy}
\end{equation}

We assume incompressibility:
\begin{equation} \label{eq:3}
     \nabla \cdot \textbf{v} = 0
\end{equation}

And for each timestep, the flow is irrotational such that for a typical closed contour not containing a singularity: 
\begin{equation} \label{eq:4}
    \nabla \times \textbf{v} = 0
\end{equation}

Another way of saying this is that there is not a net vorticity, (unless it is constant and generated by a singularity in the flow, which constitutes an exception).
\newline
\newline
Equations \ref{eq:3} and \ref{eq:4} lead to Poisson's equation for the velocity potential: 

\begin{equation} \label{eq:5}
    \nabla^2 \phi= 0
\end{equation}

Velocity potentials ($\phi$) are solutions that satisfy the Poisson equation \ref{eq:5} and obey continuity \ref{eq:3}. These solutions are linearly superpositionable.
\newline
\newline
Equation 5 becomes in 2D polar coordinates:
\begin{equation} \label{eq:6}
    \frac{1}{r} \frac{d}{dr} (r \frac{d \phi}{dr}) + \frac{1}{r^2}\frac{d^2\phi}{d\theta^2} = 0
\end{equation}
\newline
\newline
where r is the radius from the origin. 
\newline
\newline
Now, say we would like to solve for the velocity field ($u,v$) that occurs because of an expanding incompressible volume (source), such as our active bubble. We will solve this in the unsteady domain such that the radius of the expanding circle is equal to a function $a(t)$ (defined later). 
\newline
\newline
We see that $\phi$ is a function of $r$ and $t$ only. 
\newline
\newline
The boundary condition will be a velocity condition: the fluid at the expanding bubble edge needs to match the driving velocity: 
\begin{equation} \label{eq:7}
    \frac{\partial \phi}{\partial r}  = \frac{da}{dt} \; \textbf{at} \; r = a(t)
\end{equation} 

Equation \ref{eq:6} then simplifies to:

\begin{equation} \label{eq:8}
    \frac{\partial}{\partial r} (r \frac{\partial \phi}{\partial r }) = 0
\end{equation}

\begin{equation} \label{eq:9}
    r \frac{\partial^2 \phi}{\partial r ^2} = - \frac{\partial \phi}{\partial r}
\end{equation}

Integrating both sides, we see:

\begin{equation} \label{eq:10}
 \int_{}^{} r \frac{\partial^2 \phi}{\partial r ^2} \; dr =  \int_{}^{} - \frac{\partial \phi}{\partial r} dr
\end{equation}

\begin{equation} \label{eq:11}
    r \frac{\partial \phi}{\partial r} - \bcancel{\phi} = -\bcancel{\phi} + c_1
\end{equation}

Apply boundary condition given in equation \ref{eq:7} to solve for $c_1(t)$:

\begin{equation} \label{eq:14}
    c_1(t) = a \frac{da}{dt}
\end{equation}

\begin{equation} \label{eq:15}
   \frac{\partial \phi}{\partial r} = \frac{a}{r} \frac{da}{dt}
\end{equation}

Integrate a second time: 
\begin{equation} \label{eq:16}
    \phi = a(t) \frac{da}{dt} \ln(r) + c_2
\end{equation}

Set $c_2 =0$ to set velocity potential relative to origin.

To solve for the radial component of the velocity, take the derivative of the velocity potential with respect to r: 

\begin{equation} \label{eq:17}
    u_r = \frac{a}{r} \frac{da}{dt}
\end{equation}

Now say the source is oscillating such that:
\begin{equation} \label{eq:18}
    a(t) = \dfrac{r_{max} - r_{min}}{2} \sin{2 \pi t} + \dfrac{r_{max} + r_{min}}{2}
\end{equation}
Where $r_{min}$ is the minimum active bubble radius, and $r_{max}$ is the maximum active bubble radius. Alternatively, (for comparison to our actual experimental data), we can define $a(t)$ as a piece-wise ramp and hold function similar to the one shown in figure 2c. 

For a known maximum radius and minimum radius of the active bubble, we can approximate the projected position of any initial coordinate in the 2D potential flow for any point in time. 
\clearpage
\section{Sample Preparation}
\label{sampleprep}
Preparation of the foams discussed in this paper were done by hand, by cleaning and then wetting a glass plate with a surfactant-laden solution. Bubbles were generated by moving a steel blunt-tip needle, partially submerged, in the wetting layer, dispersing nitrogen gas at a gauge pressure of about 0.75 psi. A second wetted glass plate was then set 1 millimeter  above the bubble layer, with spacers controlling the gap distance between plates. The resultant foams have the properties that they are polydisperse (see example distributions in Appendix B) and single-layer (2D), with volume fractions under ten percent. Because of our preparation technique, these are frustrated structures far from any global energy minima (unlike hyperuniform foams which have experienced significant annealing through long-term relaxation and coarsening  (Supp Ref. 2). Our initial foams are internally diffusive in character, and have potential structural energy that relaxes over very long timescales (order 10-30 hours), in the absence of other perturbations. The preparation of the initial foam introduces quenched disorder in the structure, that in the absence of perturbation, only relaxes on timescales far longer than that of the experiment.

\section{Properties of Initial Foam Structures}
\subsection{Quenched disorder and Effective Temperature}
Because of the way in which these foams are made (see: section \ref{sampleprep}), the structure can be quite fragile and soft. The placement of the top plate over the bubble raft forces the geometry into a new, more frustrated configuration over a short period of time (much like "quenching"). This locks in a fragile, disordered state that is mechanically quite irregular.

\begin{figure}[ht!]
\centering
\includegraphics[width=\linewidth]{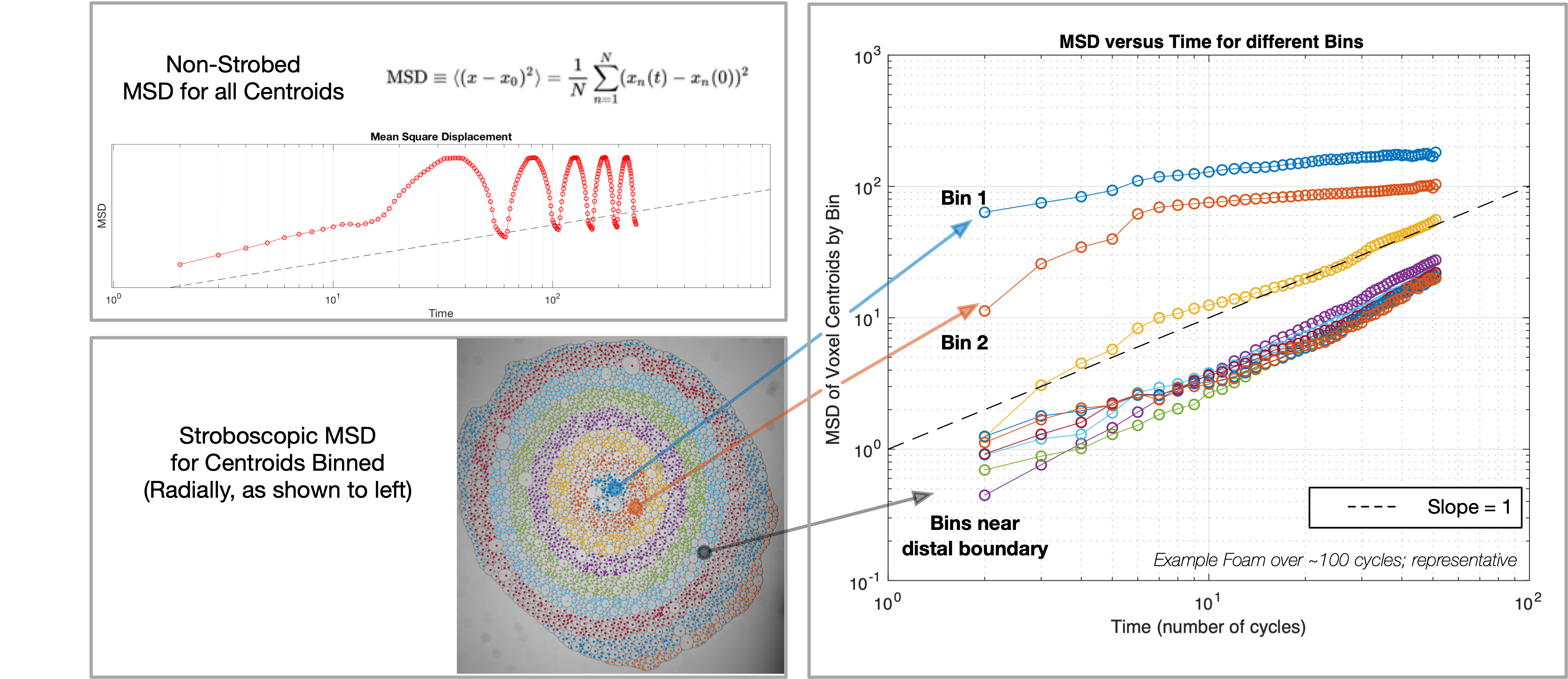}
\caption{Top Left: Mean Squared displacement as a function of time is shown. Bottom Left: We can define bins, moving radially outward from the site of activity. Right: Mean Squared Displacement is shown for each bin, where the x-axis is sampled stroboscopically (only includes the first frame of each cycle). This highlights long-term adaptation of the structure through many cycles.}
\label{MSDPlot}
\end{figure}

Although there is very little true noise / or thermally-induced fluctuations in our system, the prepared foams in this study are interestingly not in a particularly ultra-low or zero temperature state. Although they are jammed (the foams  prior to an experiment are not undergoing spontaneous T1s or exhibiting any directed motion over timescales on the order of many minutes), the centroids of the cells move as if they had an effectively diffusive nature. The evidence for this is that the mean squared displacement of the centroids of the cells prior to a perturbation has a diffusive character, characterized by the slope of approximately one on the log-log MSD plot in figure \ref{MSDPlot}.

These foams are highly frustrated and the vertex motion is “fluid-like”, despite their volume fractions under 10 percent, and the confluent, globally jammed nature of the bubbles. The initial structure is storing energy, and the internal motion of this structure dissipates this energy very slowly (via shear on lubrication films at the top and bottom plate), as it moves down a highly frustrated landscape. 

\begin{figure}[ht!]
\centering
\includegraphics[width=4in]{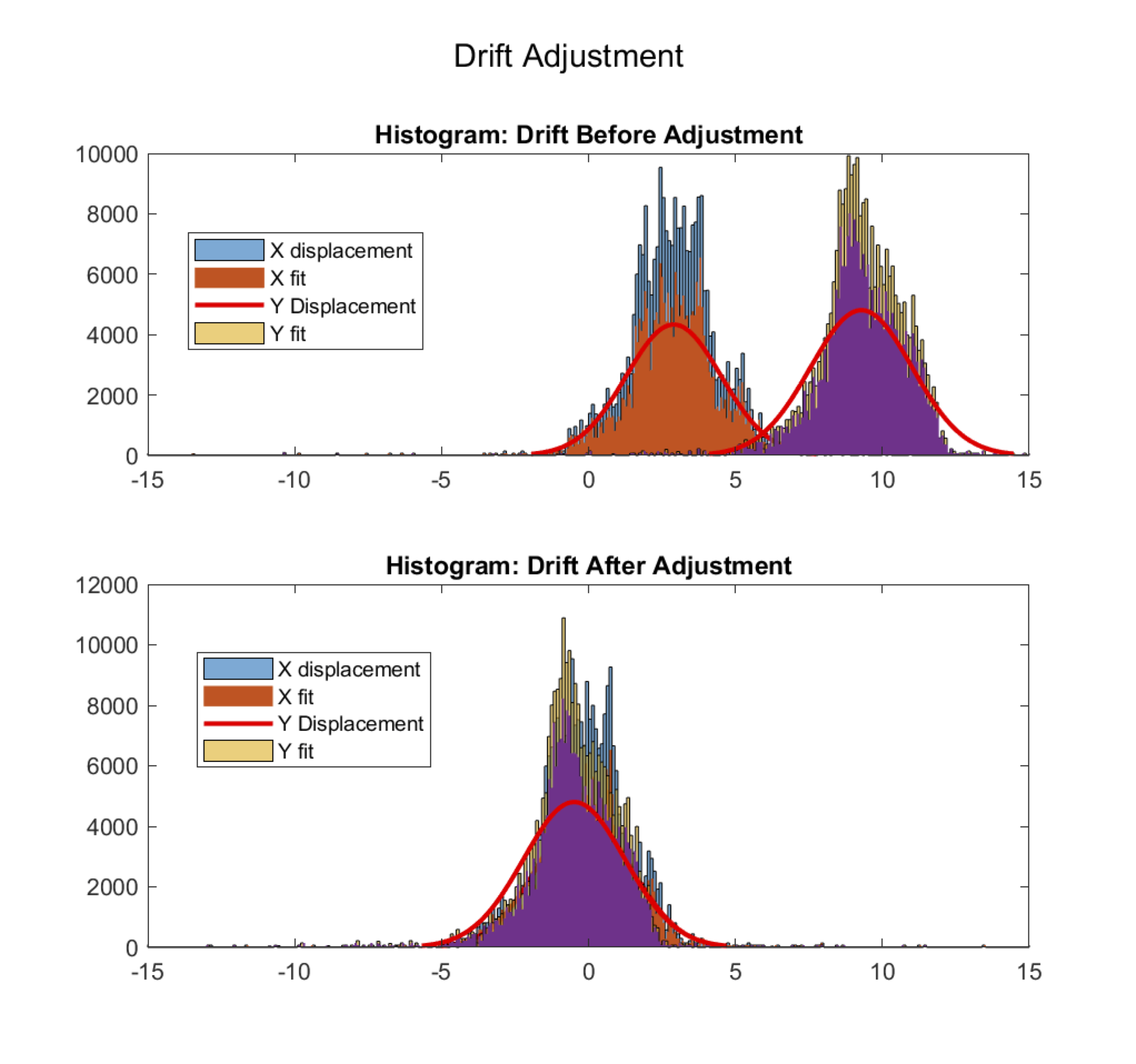}
\caption{Prior to computing MSD, it is  necessary to correct the displacements first for any background drift (caused by gravity / occasional error in the flatness of the experiment) as shown in this plot}
\label{fig}
\end{figure}

\newpage
\clearpage

\newpage
\clearpage
\subsection{Polydispersity and Coordination Number}
\label{polydispersityHistogram}
Histograms are shown here for the polydispersity (area distributions) for the 3 videos (15 cycles) used for the experimental analysis in Figures \ref{experimentalPlatform} through \ref{energy}. As expected, the median and average coordination is 6 (a robust topological feature of foam).

\begin{figure}[ht!]
\centering
\includegraphics[width=\linewidth]{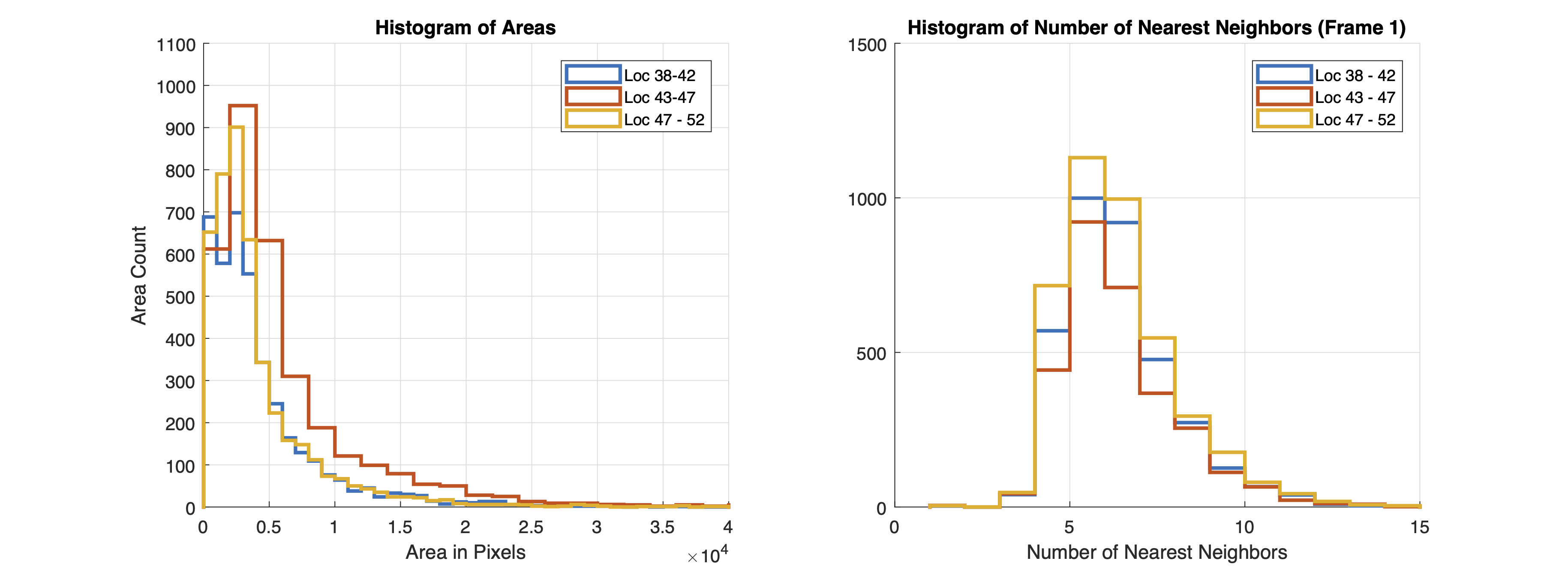}
\caption{Polydisersity histogram of the 3 example datasets discussed in paper}
\label{fig}
\end{figure}
\newpage
\clearpage

\subsection{Edge Dispersity and distance to yield}

\begin{figure}[ht!]
\includegraphics[width=\linewidth]{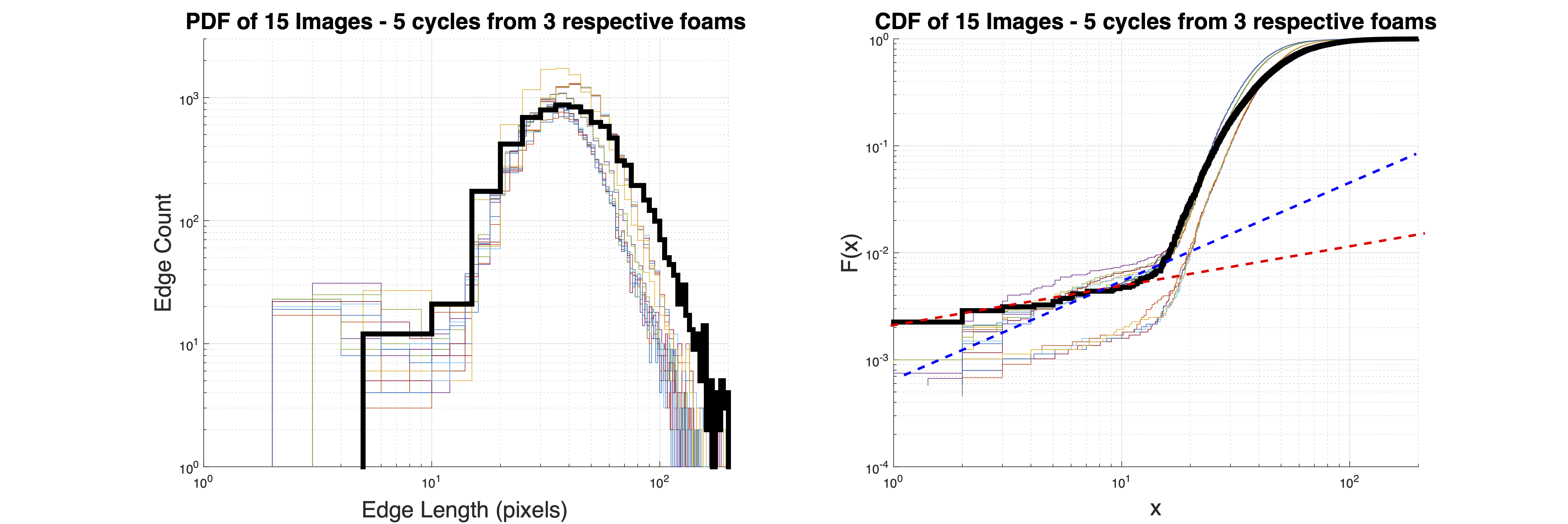}
\caption{Edge Dispersity PDF and CDF of the first 5 cycles of each of the 3 example datasets discussed in paper}
\label{fig}
\end{figure}



\section{Simulation for Micro-structural Simulation of Vertex Motion}
\label{simulationDetails}
In an effort to understand the physical, mechanical origins of the phenomenological ‘loop’ nature of the vertex trajectories close to the site of injection, we have developed a simple model that demonstrates that these trajectories are breaking symmetry fundamentally due to the 3-fold symmetry of the microstructure, and a discrepancy in relaxation timescales between vertices and edges.  

Our reduced order model takes several approximations. The first of which is that each edge in the system is, at all times, a circular arc segment. The arc segment is fully defined by a single degree of freedom, (defined as $h$), the height of the arc segment relative to the chord at the center-point between the two vertices. In addition to each edge having a single degree of freedom, each vertex has 2 degrees of freedom (position in x and y). For the simple model system, we studied a single mobile vertex attached to 3 edges — a system of five 2nd order, coupled nonlinear ODEs (3 edges each with a single degree ($h_{1 - 3}$), 1 vertex with two spatial degrees (in $x$ and $y$).

Each coupled ODE is given by a simple force balance which naturally give rise to the familiar equilibrium laws such as Plateau’s law and Laplace’s law.

\begin{figure}[ht!]
\includegraphics[width=\linewidth]{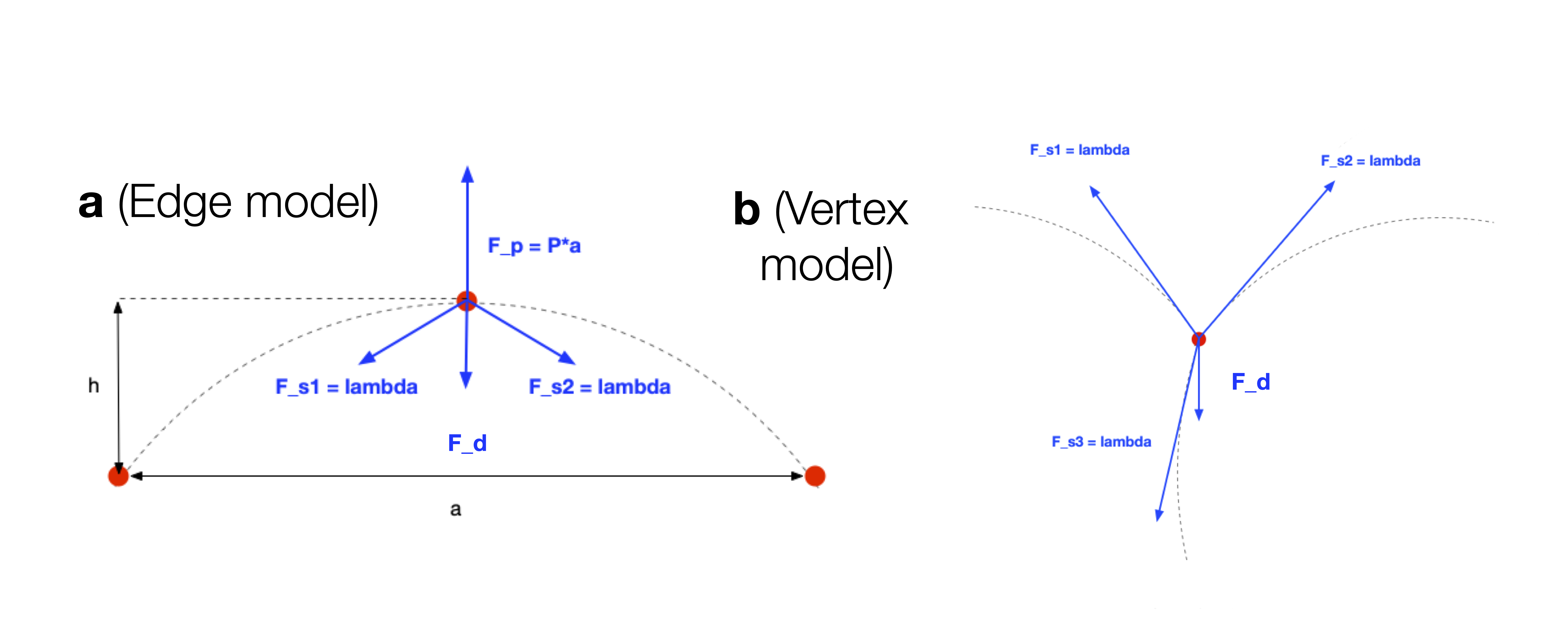}
\caption{Free body diagrams for the force balance at of the edge objects versus the vertex objects}
\label{freebodydiagram}
\end{figure}

The force balance on the vertex is given by 
\begin{equation}
    F_{vertex} = F_1+F_2+F_3+F_D
\end{equation}

Where $F_1$,$F_2$ and $F_3$ are  forces with a magnitude of $\lambda$ (surface tension), and a direction exactly tangent to the edge at the position of the vertex. The orientation of these three forces defined by the edge curvature through this tangent constraint, coupling the vertex motion to the edge motion. 

The force balance on the edge node in the center of the edge is given by:
\begin{equation}
    F_{edge} = P_{Diff} c + F_D
\end{equation}

Where $P_{Diff}$ is the pressure difference across the edge, $C$ is the chord length between the edges, and $F_D$ is a nonlinear damping term, proportional to the velocity to the $2/3$ power.

The boundaries of the system are fixed / pinned in this model, such that each edge has a distal vertex that cannot move its position. 


In this model we can study the motion of the mobile vertex in response to a step function in pressure of the 3 surrounding bubbles. 

The total force on each node (given in equations 24 and 25) is used to simultaneously solve the second-order system numerically, using a Runge-Kutta integration method. 

The purpose of building an inertial solver is to clearly demonstrate the mechanics of the problem in terms of known underlying mechanisms that drive micro-structural dynamics (e.g., pressure, surface tension). We chose not to use a gradient-decent analog of this model, specifically because it slightly obfuscates the underlying forces involved by enforcing a cost function based on volume conservation. However, it is noted that an energy-based minimization method should also work to demonstrate these phenomena, if the motion of the vertices and the edges have different relaxation timescales. In other words, the looping and non-closure phenomena we demonstrate with this model do not appear to be inherently inertial in origin, as the behavior is extremely robust to large changes in the "mass" terms.  

\begin{figure}[ht!]
\centering
\includegraphics[width=0.5\linewidth]{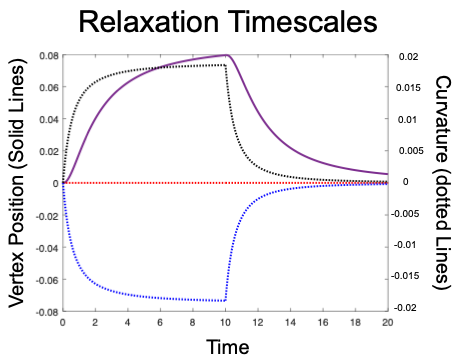}
\caption{Simulations of three-edge model demonstrate that the edges have faster relaxation time than vertecies in addition to a subtle phase lead.}
\label{fig}
\end{figure}







\section{Multi-bubble Simulation Methods}
We extend the dynamical simulations of edge and vertex dynamics to a more realistic system of bubbles. 

\subsection{Galilean Transform on Edge Motion Equations}
When we study edges in a foam, each edge's motion is complicated by the fact that the entire edge may be moving significantly (moving reference frame because of the fact that the two vertecies are translating). Before extending the edge model to full foam systems (as discussed in figure \ref{chirality}h-g), we must be careful to account for the motion of the reference frame of the edge. 

This is accomplished with a Galilean transform correction on the acceleration of the edge in the direction normal to the chord of the edge $\vec F_{edge}$:


\begin{equation}
   \frac{d^2h}{dt^2} = \frac{1}{m_{\text{edge}}}((\vec P_{Diff}) c+ \vec F_{\lambda}+ \vec F_{D,edge}) + \frac{dV_{\text{frame,}\hat{k}}}{dt}
\end{equation}

The acceleration of the edge height, $\frac{d^2h}{dt^2}$, is corrected by a small term to account for the reference frame acceleration in the $\hat{k}$ direction (perpendicular to $c$, the chord of the edge). This correction term uses the current acceleration of the vertex nodes, computed earlier within the same timestep. This reference frame correction is a common issue when dealing with height-function treatments on free-boundaries (Supp. Ref. 4). 

The limitations of imposing this reference frame correction is that errors will accumulate in the edge shape approximation if significant rotations and or significant shear forces are applied to the edge. For the modelling of foam in response to predominantly normal forces (inflation and deflation; tension or compression) these approximations are adequate. 

Thus, this model would perform relatively poorly in modelling foam flows with significant shear forces (which are often visible in the foam by the observation of non-circular arc edge segments). In such cases, one is advised to consider using other conventional methods (Supp refs 1,6) which use fewer approximations. 

\subsection{Bubble pressure modelled as an ideal gas}
Bubble pressures were modelled as an ideal gas: 

\begin{equation}
    P_iV_i^\gamma = n_iRT
\end{equation}

Where at each timestep, the pressure in each bubble $P_i$ is computed from the volume of the bubble $V_i$. At the start of each simulation, $n_iRT$ is given (e.g. assigning a fixed number of gas molecules to each bubble). It is assumed gas molecules do not migrate across films in this model.  The volume is computed at each timestep from the $x, y$ position of the vertecies and the height of the edges. $\gamma$ is the adiabatic exponent which in these simulations is set to $1$. One could vary the gas compressibility by modifying this exponent.   

\begin{figure}[ht!]
\includegraphics[width=\linewidth]{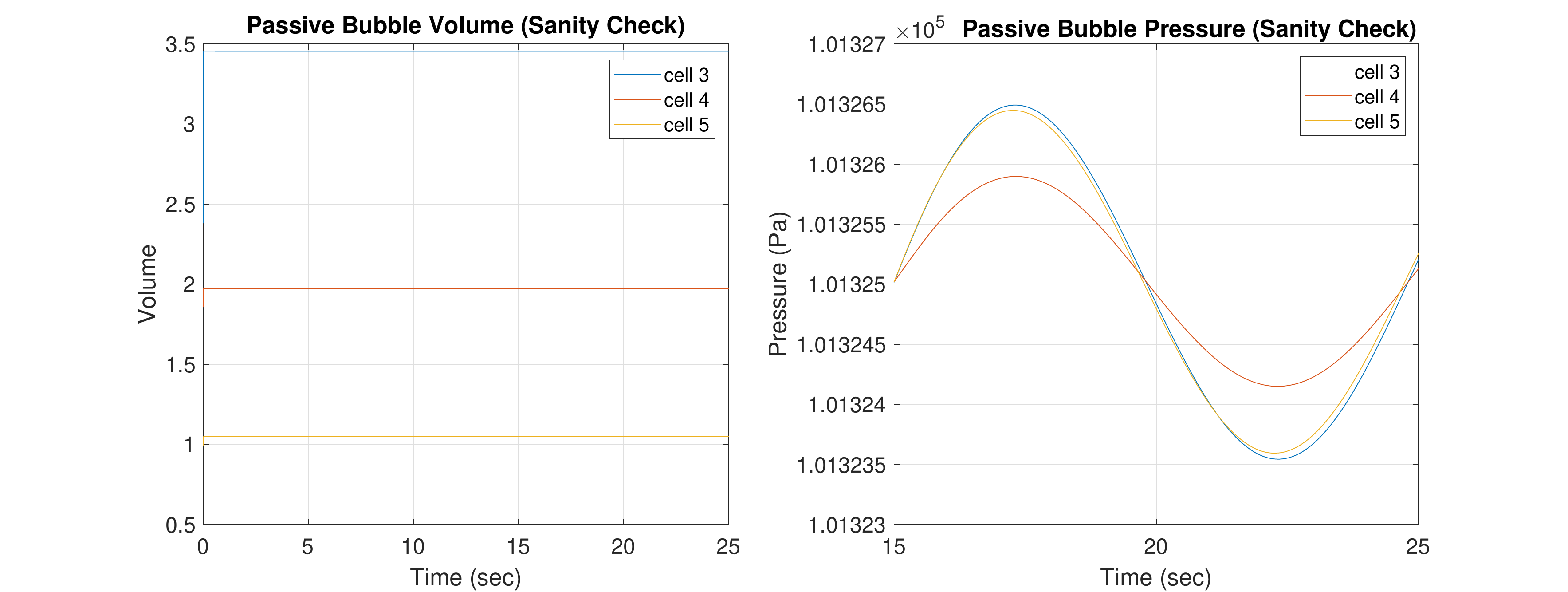}
\caption{Simulation Method Sanity Check: Peak-to-trough maximum variation in pressure on the passive bubbles is approximately $3*10^{-4}$ psi. Volume variation is $2.8*10^{-5}$ for $deltaV/intialV$ or stated differently, peak-to-trough volume change is is only $0.003$ percent of the initial original volume.}
\label{sanity}
\end{figure}

Peak-to-trough maximum variation in pressure on the passive bubbles is approximately $3*10^{-4}$ psi. Volume variation is $2.8*10^{-5}$ for $\frac{deltaV}{intialV}$ or stated differently, peak-to-trough volume change is is only 0.003 percent of the initial original volume. This is shown in figure \ref{sanity}.  

The point here is to note that although we treat the bubbles as an ideal gas, the net volume variation within a bubble is exceptionally small. Because we are working with only polygons and circular-arc segments, the computation of bubble volumes is an analytical function not dependent on a mesh or grid size. This gives us the volume and pressure in an analytical form (independent of a mesh size). 

On benefit to such a model is that more complex processes that are highly dependent on these subtle changes in pressures could be easily added to this model. For example, because at every step we track the volume and pressure of each bubble, it would be an simple extension to add the dynamics of coarsening to such a model -- where $n_i$ for each bubble is a function of time based on the pressure differences and length of the edges. 

One could also represent rough empirical behaviors of interior liquid flows in the foam films by allowing the mass terms of the edges of vertecies to be functions of time, governed by the properties of the fluid. Such extensions are feasible additions to this framework. 


\section{Spring Network Models}
\label{springs}
\begin{figure}[ht!]
\includegraphics[width=\linewidth]{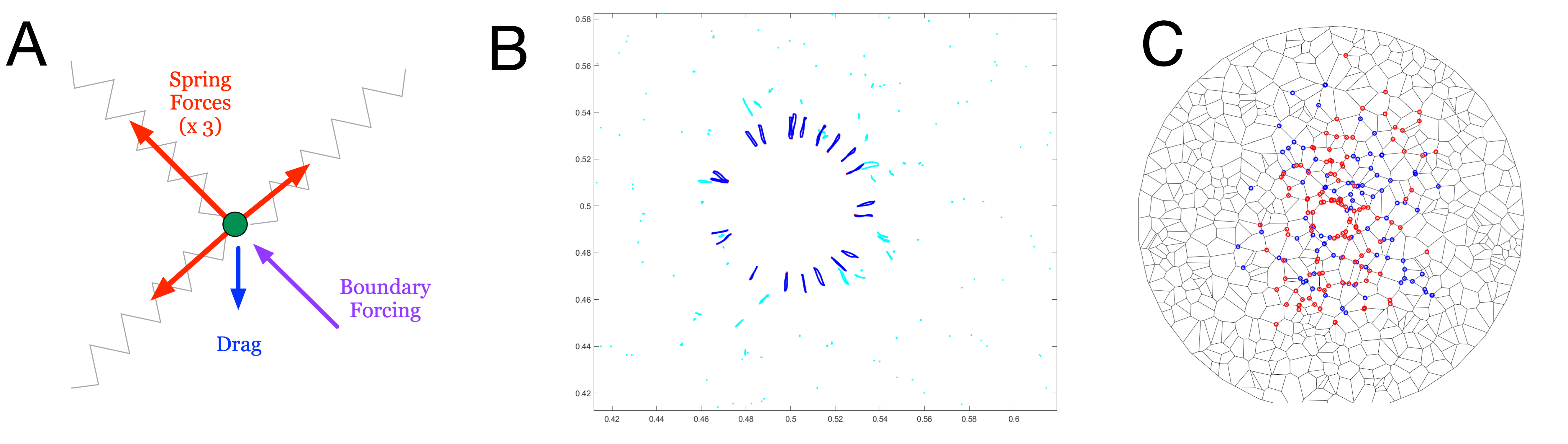}
\caption{A simple spring model network demonstrating very minimal local cooperation in an inflating and deflating network. In this system, cells do not conserve volume, which screens the inflation and deflation very locally to the active site. }
\label{springFig}
\end{figure}

As shown in Figure \ref{springFig}, we can apply the same lock-in amplifier method (SI \ref{lockin}) to a simple linear spring network to compare with foam-like systems. We constructed a network where force balances on the nodes are given by the attached springs and a drag force (Fig \ref{springFig}a). The springs varied in length, but all had the same linear spring constant, $k$. An additional boundary force was applied to those nodes adjacent to the active bubble. 

The resulting displacement of the network in response to a single inflation-deflation is shown in Fig \ref{springFig}b. 

The lock-in amplifier technique is shown in figure \ref{springFig}c, where red dots correspond to nodes with a CW swirl and blue nodes correspond to nodes with a CCW swirl. We do not observe coherent large-scale CW/CCW in these linear systems, unlike the foam we study in section \ref{chiralitySection}.   

\section{Methods for Segmentation of Experimental Data}
We have developed a set of segmentation and tracking methods for these cellular systems, which we provide upon reasonable request. These image-processing methods are relatively standard in the study of foams and emulsions, and centrally utilize the MATLAB built-in "Watershed" algorithm, which uses the Fernand-Meyer method (Supp Ref. 8).   

\section{Measuring CW/CCW "Swirl" Directionality in Vertex Trajectories}
To measure the direction of swirl of the vertex trajectories, we implemented a lock-in amplifier technique as described in detail below. 

\subsection{Lock-In Amplifier for Binary Directional Measurement}
\label{lockin}
By treating the y-axis as an imaginary parameter and the x-axis as a real parameter, then we can represent point motion using a complex variable, 
\begin{equation}
    P_{exp} = K e^{x},
\end{equation}
that varies as a function of time. P is the position of the point, K is an amplitude variable which is considered strictly positive, t is time, and x is a variable with the form of
\begin{equation}
   x = i (\omega_{exp} t + \phi).
\end{equation}
$\omega_{exp}$ is either approximately $-\omega$, representing clockwise motion, or $+\omega$, representing counterclockwise motion. The motion of the point also starts off in an unknown orientation which can be taken into account by including the variable $\phi$, as a phase offset for the experimental data. In order to test the hypothesis of whether the experimental variable is is closer to $-\omega$ or $+\omega$ we can create two comparison cases, one for clockwise motion,

\begin{equation}
    P_{CW} = e^{-i \omega t}
\end{equation}

and a similar comparison for counterclockwise motion,
\begin{equation}
    P_{CCW} = e^{i \omega t}.
\end{equation}

Comparison of the hypothetical cases with the experimental case can be accomplished by dividing the experimental motion variable with the reference variables and integrating over a full cycle, giving two cases, 

\begin{equation}
   A = \int_{0}^{\frac{2\pi}{\omega}} \frac{P_{exp}}{P_{CW}} dt = \int_{0}^{\frac{2\pi}{\omega}} K e^{i \omega_{exp} t + i \omega t + i \phi} dt
\end{equation}
and the counter clockwise test of
\begin{equation}
   B = \int_{0}^{\frac{2\pi}{\omega}} \frac{P_{exp}}{P_{CCW}} dt = \int_{0}^{\frac{2\pi}{\omega}} K e^{i \omega_{exp} t - i \omega t + i \phi} dt.
\end{equation}

In order to determine which hypothesis is more valid, we can take the ratio of A and B
\begin{equation}
   C = \frac{A}{B} =  (\frac{e^{i 2 \pi (\omega_{exp}+\omega)}-1}{\omega_{exp}+\omega})*(\frac{\omega_{exp}-\omega}{e^{i 2 \pi (\omega_{exp}-\omega)}-1})  = \frac{\omega_{exp}-\omega}{\omega_{exp}+\omega}
\end{equation}
which allows us to cancel out the factor of $\phi$ and the amplitude parameter K from the experimental data, allowing the system to become wholly dependant on the comparison frequency, $\omega$ and the experimental rotation frequency $\omega_{exp}$. If C is greater than 1 then it indicates that hypothesis A proved a better fit than hypothesis B, and the point moves in a clockwise direction. A value for C less than 1, conversely indicates motion in a counterclockwise direction. While this method is simple, it provides a robust framework for detecting the time based motion of the directionality of points in spite of noise and the reality that K varies throughout the motion of the trajectory.

\section{Cellular Segmentation (based on Watershed algorithm)}
\label{segmentation}
For each dataset video we can extract all possible information about the structure in each frame: 
\begin{itemize}
\item the voxels (or cells),
\item the location and area of each cell, 
\item the connectivity of which cells are neighbors, 
\item the events flagged for when and where T1 events occur, 
\item the vertex locations throughout the video, 
\item the vertex connectivity, 
\item the edges (in 3 formats: straight-line approximations, binary data representing the real edge shape curve, and circular arc fits, reducing each edge to only a length and a curvature DOF). 
\end{itemize}
Below we provide an overview of each sequential step in the process. This is also depicted visually in SI Movie 2.

The first step is removing background noise and gradients from the image with a series of background-subtraction steps. This step may be unnecessary for extremely clean data with exceptionally flat illumination. 

The second step is using the watershed algorithm to segment each image into cellular regions. $$\text{watOut}(\text{frameNumber}) == 1 $$ contains a binary skeleton of the foam which is exactly 1 pixel wide (regardless of the foam volume fraction). $$\text{watOut}(\text{frameNumber}) = \text{localID}$$ corresponds to the image region associated with a voxel, named by that local ID. Properties of the voxel are extracted from each of these subregions.

\subsection{Network Connectivity}
The centroid of each cell is computed and we use a region adjacency method (Supp Ref 7) to identify which cells are neighbors. This results in a connectivity matrix, describing which cells are adjacent (with “local” ID numbers that describe the cell in a frame).

A simple tracking algorithm (Supp Ref 3) is then used to associate the local IDs of each cell in a given frame to a global ID for the cell across all frames in a video, such that each individual voxel has a single ID number in a video. Each cell is saved as a voxel object, with properties that record its centroid location and time series of local IDs over the video. Segmentation errors will sometimes cause the voxel to persist for less than the entire video; typically this is due to tracking assigning the voxel a new global ID at some point in the video. This error may be tuned using fitting parameters in track.m. 

The area of each cell is computed and saved. 

A connectivity tensor is built (from the set of local connectivity matrices), that associates the connectivity of the global voxel IDs in the entire video.

\begin{figure}[ht!]
\includegraphics[width=\linewidth]{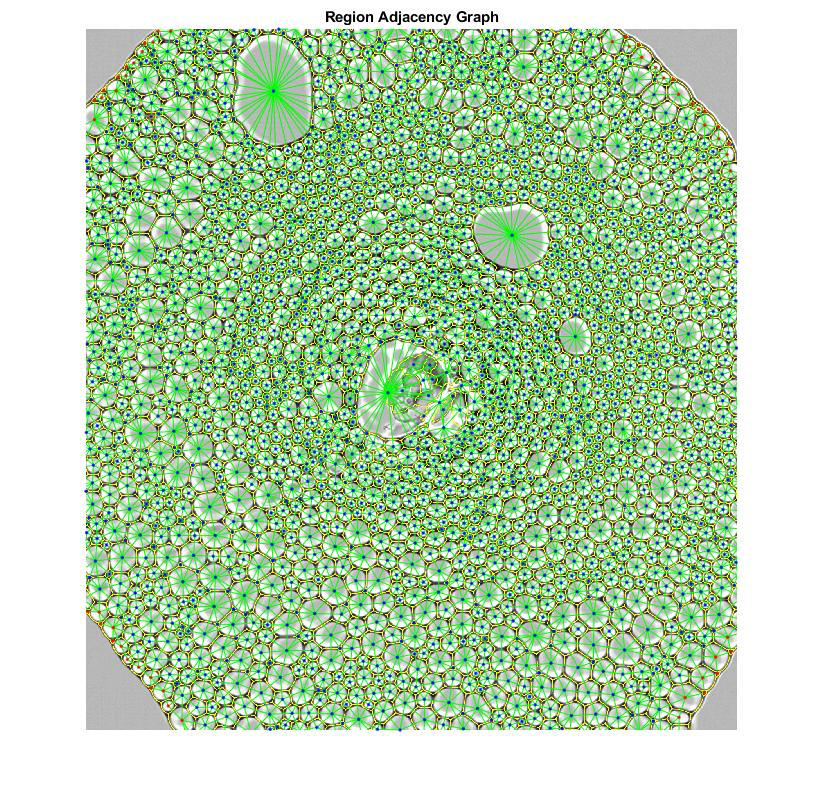}
\caption{Adjacency plot superposed over original foam image (this is an abnormally wet foam, not one from the main paper datasets). Green lines represent connections between neighboring voxels.}
\label{fig}
\end{figure}

\subsection{Vertex Identification}
Further, we track the location of each vertex in the foam. This is done by using the skeleton from the segmentation step. Because the watershed algorithm always produces a skeleton that is exactly 1 pixel in width in all places, the vertex locations have a particular signature 3x3 pattern. A 3x3 by 50 tensor of all possible vertex patterns is built. The binary skeleton of each frame is convolved with each 3x3 pattern in a loop (this is done in parallel for faster compute time), and matches to the pattern are exact. Error is exceptionally low and typically occurs only in exact locations of T1 events (where 4 edges meet). This is a potential method to perform additional checks on the T1 Identification methods previously described. 

The vertexes are then associated with each voxel object in each frame. For each voxel region, the boundary is computationally grown or expanded by a small amount (tunable, but typically only 1 pixel is necessary). Vertexes within the boundary are associated with that voxel. This is repeated for all voxels in each frame.

Vertices are then tracked throughout the video using the same tracking algorithm as for the centroids, to assign a persistent identity to each junction.

\begin{figure}[ht!]
\includegraphics[width=4in]{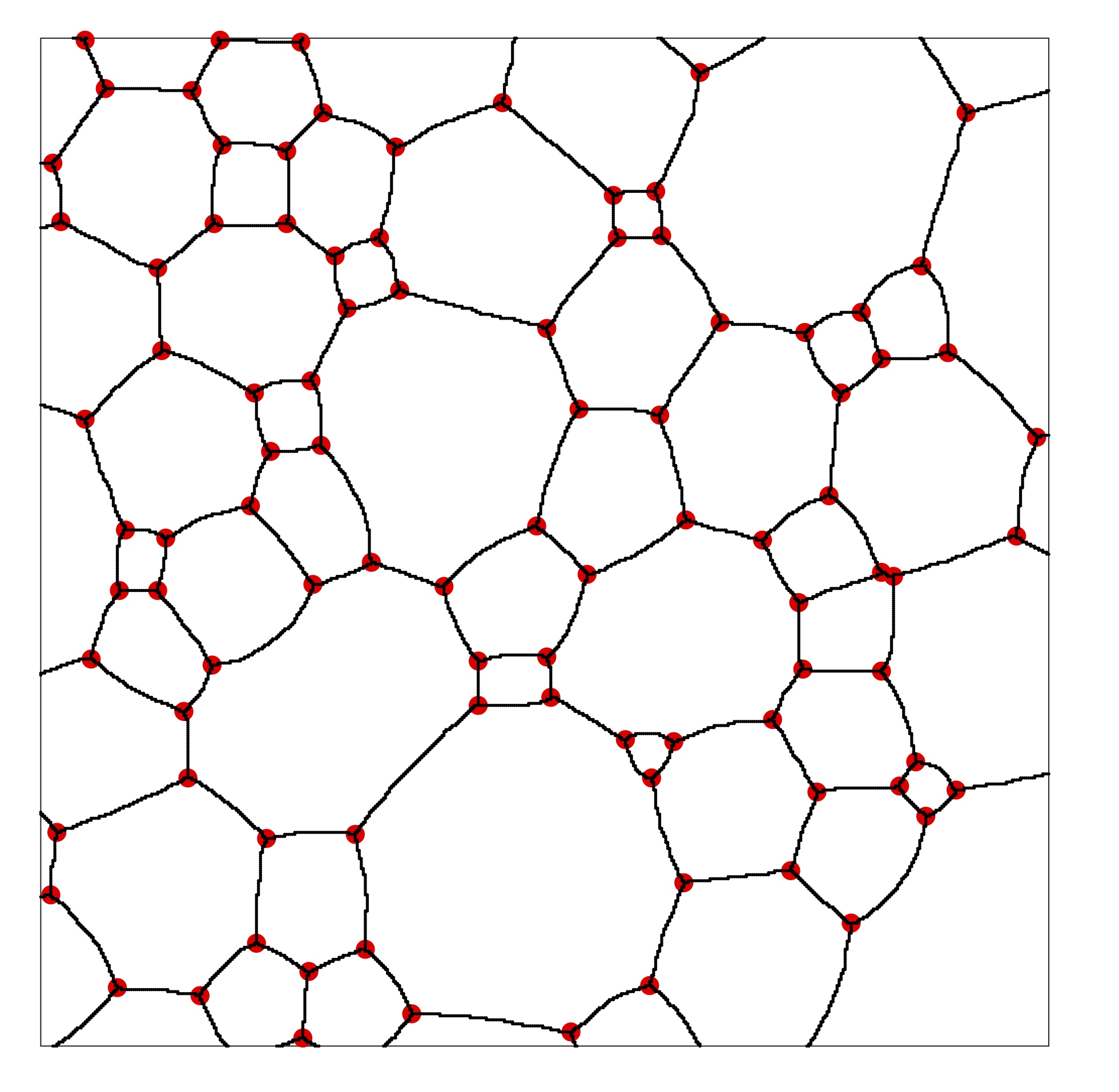}
\caption{Vertex Finder demonstrates high accuracy in identifying vertex positions (red)}
\label{fig}
\end{figure}

\subsection{Edge Identification}
To compute connectivity of the vertices (to achieve data on edge length distributions), the vertexes associated with each cell are connected counter-clockwise for each voxel object, populating a connectivity matrix for the vertices. This connectivity matrix is 3-regular , and has an interesting mathematical relationship to the connectivity matrix of cellular centers (which is irregular). 

This method is redundant (each edge is counted twice), however the result is exceptionally accurate (no missed edge has yet been identified). Each edge is approximated by a straight line between the voxels — however, we also have the binary edge data (which contains all curvature information as shown in Fig. 17) in the original skeleton. Straight-line edge data is shown in supplementary figure 16. Early algorithms are under development that have used the binary edge and the straight-line abstracted edge that connects the two vertexes to compute curvature for each edge, based on a simple circular-arc fitting method. We then use these circular arcs as edge models to compute  the 3 internal angles at each of the vertices.

\section{Methods for T1 Transition Identification}
\label{T1Identify}
This voxel connectivity tensor is analyzed for particular local connectivity changes that are unique to a T1 event. A series of initially detected events are cataloged. However, due to known errors in the region adjacency method (to improve speed of the algorithm), a filtering step is necessary to identify which events are truly T1 events versus connectivity events that arise from errors in the segmentation and adjacency methods. This filtering checks that each cell in an event is persistent for a certain minimum time before and after the event, and that the connectivity in the region obeys certain continuities. 

A final list of T1 events is produced that contains the location of the event, the frame number in which it occurs, and the global ID numbers of each of the cells that was involved. 

\begin{figure}[ht!]
\includegraphics[width=\linewidth]{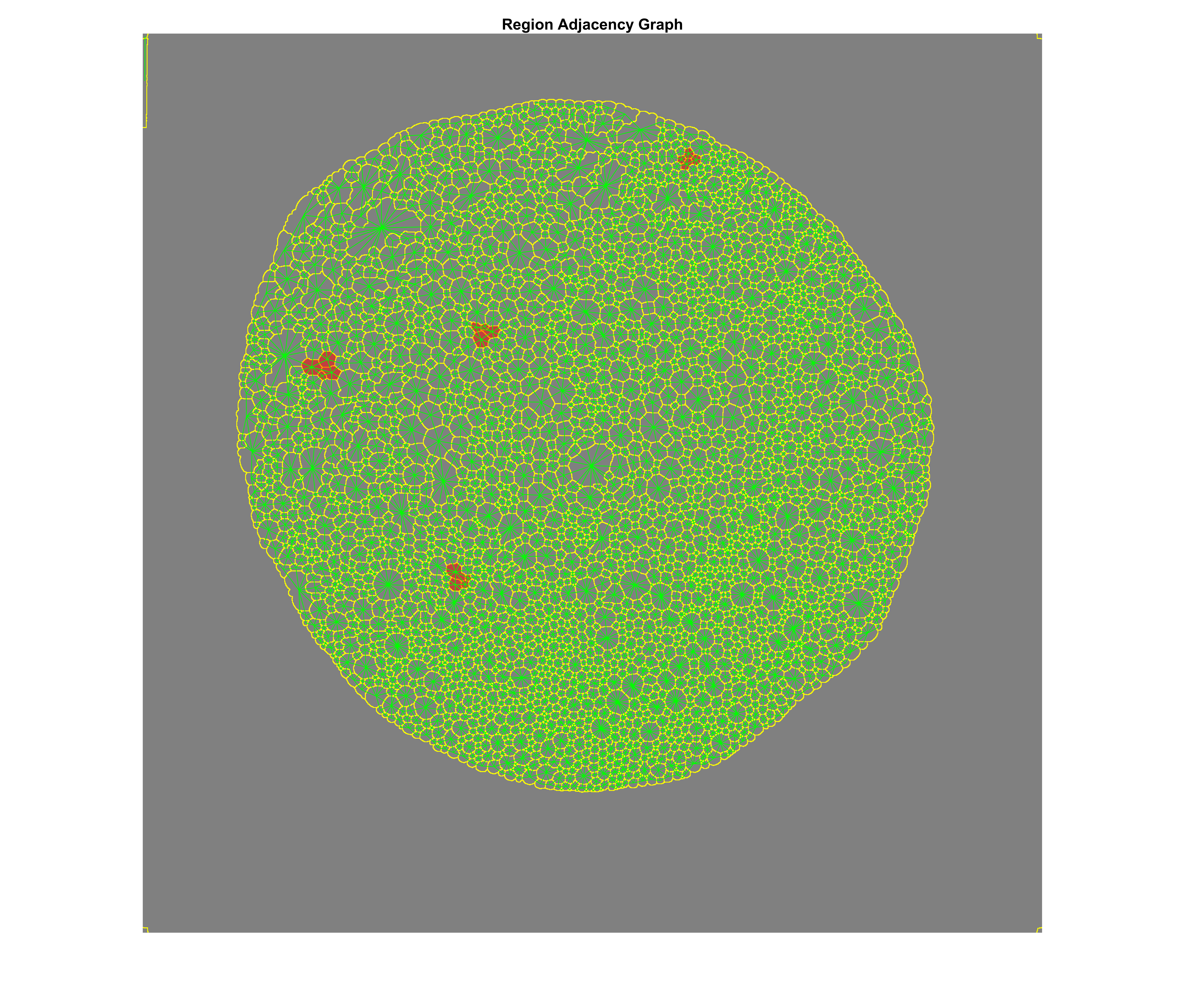}
\caption{Adjacency plot superposed over original foam image (this is an abnormally wet foam, not one from the main paper datasets). Green lines represent connections between neighboring voxels.}
\label{fig}
\end{figure}

\section{Data and Code Availability}
Raw Data and segmented data will be available here (upon peer review): \url{https://tinyurl.com/2eey79up}
Simulation Data will be availible here (upon peer review): \url{https://tinyurl.com/4duj2hp6}
Code for segmentation and experimental data feature will be available on Github upon peer review.
Simulation code for the dynamical model and statistical model may be provided upon reasonable request. 

\clearpage
\newpage
\includepdf{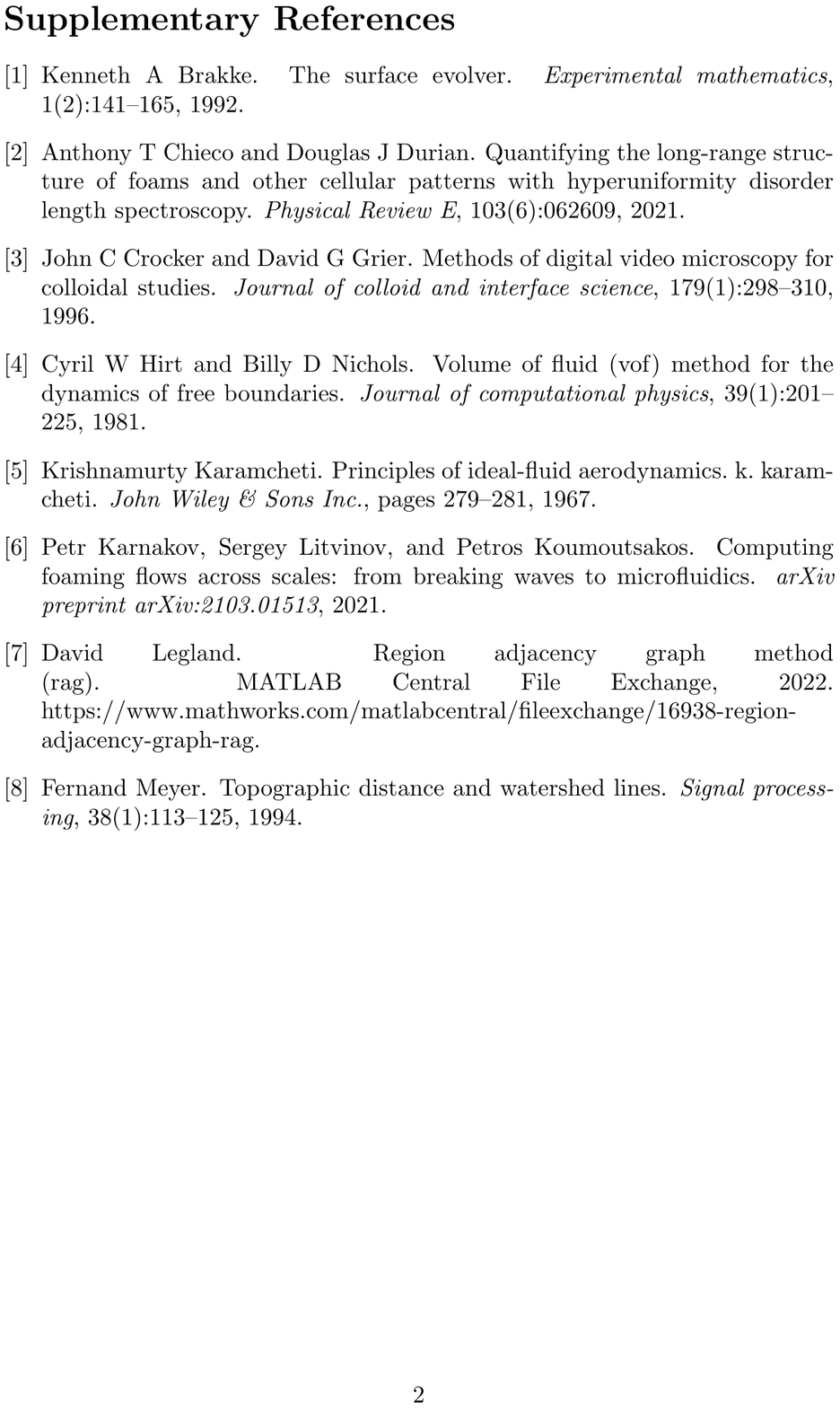}

\end{document}